\documentclass{emulateapj-rev}

\usepackage{epstopdf}
\usepackage{graphicx}
\usepackage{natbib}
\usepackage{amsmath}

\def\msun{\ifmmode {\rm M}_{\mathord\odot}\else $M_{\mathord\odot}$\fi}
\newcommand{\FLASH}{{\sc flash} }
\newcommand{\PARAMESH}{{\sc paramesh} }
\newcommand{\disperse}{{\sc DisPerSE} }

\newcommand{\Mcrit}{$M_{\textrm{line}}^{\textrm{crit}}$}

\newcommand{\Msol}{M$_{\odot}$}
\begin{document}

\title{The Role of Turbulence and Magnetic Fields in Simulated Filamentary Structure}

\author{Helen Kirk\altaffilmark{*,1,2,3}, 
Mikhail Klassen\altaffilmark{2},
Ralph Pudritz\altaffilmark{1,2},
Samantha Pillsworth\altaffilmark{2,4} }
\altaffiltext{*}{Banting Fellow}
\altaffiltext{1}{Origins Institute, McMaster University, Hamilton, ON, 
L8S 4M1, Canada}
\altaffiltext{2}{Department of Physics and Astronomy, McMaster University,
Hamilton, ON, L8S 4M1, Canada}
\altaffiltext{3}{National Research Council Canada, Herzberg Astronomy and Astrophysics,
Victoria, BC, V9E 2E7, Canada; helen.kirk@nrc-cnrc.gc.ca}
\altaffiltext{4}{Saint Mary's University, Department of Astronomy
and Physics, Halifax, NS, B3H 3C3, Canada}

\begin{abstract}
We use numerical simulations of turbulent cluster-forming regions 
to study the
nature of dense filamentary structures in star formation.  Using
four hydrodynamic and magnetohydrodynamic simulations 
chosen to match observations,
we identify filaments in the resulting column density
maps 
and 
analyze their properties.  We calculate the radial column density 
profiles of the filaments every 0.05~Myr 
and fit the profiles
with 
the modified isothermal and pressure confined
isothermal cylinder models, finding reasonable fits for either model.
The filaments formed in the simulations
have similar radial column density profiles 
to those 
observed.
Magnetic fields provide additional
pressure support to the filaments, making `puffier' filaments
less prone to fragmentation than
in the pure hydrodynamic case, which continue to condense at a slower
rate.  In the higher density simulations, 
the filaments grow faster through 
the increased importance of gravity.  Not all of the filaments 
identified in the simulations will evolve to form stars: some expand
and disperse.  
Given these different filament evolutionary paths, 
the trends in bulk filament width as a function of time, magnetic
field strength, or density, are weak, and all cases are
reasonably consistent with the finding of a constant filament
width in different star-forming regions.  In the simulations,
the mean FWHM lies between 0.06 and 0.26~pc for all times and initial
conditions, with most lying between 0.1 to 0.15~pc; the range in FWHMs
are, however, larger than seen in typical {\it Herschel} analyses. 
Finally, the filaments display
a wealth of substructure similar to the recent discovery of filament bundles
in Taurus.
\end{abstract}

\section{Introduction}

Filaments appear to be an important ingredient in the formation of stars.
While filaments have been known to be associated with star forming regions
for decades \citep[e.g.,][]{Schneider79,Bally87}, observations from the 
{\it Herschel Space Telescope}, particularly the Gould Belt \citep{Andre10}
and HOBYS \citep{Motte10} Legacy Surveys have underlined the prevalence
of filamentary structures within star forming regions.  With {\it Herschel}'s
unprecedented ability to sensitively map large areas of the sky,
several common properties of filaments have now been identified.
First, filaments appear to not be well represented by the
\citet{Ostriker64} equilibrium model of an isothermal cylinder;
the column density profile is shallower \citep[e.g.,][]{Arzoumanian11}.
This may indicate that magnetic fields \citep[e.g.,][]{FiegePudritz00a} 
contribute to 
supporting the filament from collapse,
although \citet{Smith14} demonstrate that filaments formed
in purely turbulent environments also have a similarly shallow slope.
Rotation may also lead to a shallower slope \citep{Recchi14}.
Second, the mass per unit length of filaments appears to correlate
with star-formation activity: filaments with mass per unit length
less than the value needed for collapse of an isothermal cylinder
\citep{Ostriker64} tend to be associated with regions which are
forming few if any stars, while filaments with supercritical
mass per unit length values tend to be associated with active
star forming regions \citep[e.g.,][]{Arzoumanian11,Hennemann12}.  
What is still unclear, however, is what forces dominate the formation 
and evolution of the filaments, and how the filaments contribute
to star formation.  For example, are the filaments formed
primarily through turbulent shocks, or under the influence
of magnetic fields or gravity?  Does turbulence control the ability of
filaments to fragment into star-forming cores?
What forces set the observed
(column) density profiles?  And do filaments primarily provide
a denser collection of gas to promote local star formation
\citep[e.g.][]{Hacar11}, or
do they play a significant role in providing a conduit of
mass for the formation of larger stellar clusters, which appear
to form preferentially at the intersection of several filaments
\citep[see e.g.,][]{Myers09,Myers11,Schneider12,Hennemann12,Kirk13}?

In this paper, we investigate the first of these issues,
namely the formation and evolution of filaments,
through the analysis of our numerical simulations.  We compare
the column density properties of filaments formed within
four different simulations: higher and lower density, and
with and without magnetic fields.  
These analyses provide a complementary look at simulations
to those recently published in \citet{Smith14}, where the 
influence of different types of turbulence on filament properties
was examined, but the effect of the inclusion of magnetic fields
or differing initial mean densities was not.

We find that while the
largest-scale structures in the gas are set by turbulent
motions, and appear similar in all four simulations,
magnetic fields and gravity do influence the properties
of individual filaments.  In particular, magnetic fields
cushion the initial turbulent gas compressions, leading to
filaments which are initially less condensed, and subsequently
evolve more slowly (due to the weaker gravitational pull) than
the corresponding hydrodynamic case.  
We note that the simulations we analyze were only able to be run for 
a few tenths of the global free-fall time, limiting our sensitivity to
later-time evolutionary trends. 
The simulated filaments
have properties which are consistent with those measured in real
filaments characterized by {\it Herschel}, suggesting that
the general insights gained with these simulations are applicable
to real molecular clouds.
Finally, turbulence and magnetic fields, 
and not just the thermal properties of molecular gas, 
appears to set the critical conditions for gravitational instability leading to 
star formation.  

In what follows, we first discuss our numerical methods and 
simulations (Section 2),  discuss the basic filament properties resulting from the 
simulations (Section 3), compare various models of filament structure (Section 4), 
and examine the effects of spatial resolution in characterizing filaments (Section 5).  
We discuss our results and their implications, as well as the limitations
of our present analysis in Section 6.

\section{Numerical Methods}

\subsection{Simulation Setup}
We used the \FLASH hydrodynamics code \citep{Fryxell2000} version 2.5 
to perform numerical simulations of molecular clumps, i.e., parsec-scale
condensations of gas capable of forming a cluster of stars. \FLASH solves 
the fluid-dynamical equations on an adaptive Eulerian grid, making use of 
the \PARAMESH library \citep{Olson+1999,MacNeice+2000}. It includes 
self-gravity, Lagrangian sink particles to represent gravitationally 
collapsing cores and (proto)stars \citep{Banerjee2009, Federrath2010}, 
and gas cooling by dust and by molecular lines \citep{Banerjee+2006}. 
Stellar properties are self-consistently evolved via a one-zone model 
\citep{Offner2009, Klassen+2012}.

We initialize our simulation volume with a turbulent velocity field. 
The turbulence is a mixture of compressive and solenoidal turbulence 
\citep{Federrath+2008,Girichidis11} with a Burgers spectrum, i.e. 
$E_\textrm{k} \propto k^{-2}$ as in \citet{Girichidis11}, and largest 
modes having a size scale roughly equal to the side length of the simulation 
box. See also \citet{Larson81,Boldyrev2002,HeyerBrunt2004}. The turbulent 
velocity field has a root-mean-square Mach number of 6.

We perform a grid of simulations in a cube-shaped volume containing either 
approximately 500 $\msun$\ or 2000 $\msun$\ of molecular gas with a power-law 
density profile scaling as $\rho(r) = \rho_c r^{-3/2}$. The choice of density 
profile is motivated by observations of dense gas associated with high-mass 
star formation \citep{Pirogov2009}; \citet{Kauffmann10} similarly analyze
a suite of dust emission and extinction maps of molecular clouds within
the solar neighbourhood, and find that those which are not forming high-mass
stars obey $\rho(r) \propto r^{-1.63}$.  
The simulation volume has a side length 
of 2~pc, and the molecular gas is at an initial temperature of 
10~K.  

These initial conditions were chosen to be representative of nearby
molecular clumps, with a focus on NGC1333, a cluster-forming region
within the Perseus molecular cloud, located roughly 250~pc away,
and currently forming a young cluster of low- and intermediate-mass stars
\citep{Walawender08}.
Using a large-scale column density map derived from 2MASS-based extinction,
\citet{Kirk06} estimate that NGC1333 contains $\sim$1000~\Msol\ within
a radius of $\sim$1~pc; the simulations contain 500 and 2000~\Msol\ within
a 2~pc cube, thus bracketing NGC1333's mean density.  
The free-fall time for these simulations is $\sim 1$ and 0.5~Myr respectively.  
A Mach number of
6 is consistent with the typical $^{13}$CO velocity dispersion measured
across NGC1333 reported in \citet{Kirk10}, and we also note is also
consistent with the standard linewidth-size relationship \citet{Larson81}.
Molecular clumps tend to have temperatures of 10-20~K \citep{BerginTafalla07}, 
and pointed observations toward dense cores in Perseus \citep{Rosolowsky08}
have a mean temperature of 11~K, although those found in NGC1333 and
other clustered environments tend to have slightly higher values
\citep{Schnee09, Foster09}.  Similarly, the dust temperature is
estimated to be slightly elevated in areas near luminous
young protostars \citep{Hatchell13}.  None of these heating effects, however,
would have been present prior to the onset of star formation in the 
region, suggesting that an initial temperature of 10~K is reasonable.

We used the same 
initial turbulent velocity field for each simulation, but compared 
magnetohydrodynamic runs with pure hydro simulations where the magnetic 
field strength was set to zero. When including magnetic effects, we 
initialize a magnetic field parallel to the z-axis with uniform field 
strength. We select a field strength for our MHD simulation so our 
mass-to-flux ratio is $\lambda \sim 1-2$; this is slightly stronger
than the typical range estimated by \citet{Crutcher2010} of $2 - 3$. 
The mass-to-flux ratio is given by
\begin{equation}\label{eqn:mass-to-flux}
\lambda = \frac{M_{tot}}{\pi R^2 <B>} \frac{\sqrt{G}}{0.13}
\end{equation}
where $M_{tot}$ is the total cloud mass, $R$ the cloud radius, and $<B>$ the
initial mean magnetic field strength.  The factor of 0.13 is required to
normalize the flux ratio relative to the critical value where the magnetic
field just prevents gravitational collapse
\citep{MouschoviasSpitzer1976,Seifried2011}. High-mass star forming 
cores typically have values $\lambda \lesssim 5$ 
\citep{Falgarone+2008,Girart+2009,Beuther+2010}.

Table \ref{table:simulation_parameters} lists the parameters for the grid 
of simulations run.
We note that while stars (sink cells) do form in all of our
simulations, as we would expect in reality, the resolution 
(50~AU) is insufficient
to correctly predict the masses of the stars that form; tests we ran with
an increased resolution led to a larger number of lower mass stars.  
This is not a problem
for our analysis, as the resolution is
more than sufficient to characterize the structure of the filamentary gas at observable
scales.  Furthermore, the simulations are stopped at an early enough time that
stellar feedback would not have had time to influence the evolution of the gas.

\begin{table*}
\tablenum{1}
\centering
\caption{Simulation parameters}
\begin{tabular*}{0.9\textwidth}{@{\extracolsep{\fill} } llccccc}
\hline\noalign{\smallskip}
\multicolumn{7}{c}{Physical simulation parameters} \\
\noalign{\smallskip}\hline\noalign{\smallskip}
Parameter & & & {\tt 500HYD} & {\tt 500MHD} & {\tt 2000HYD} & {\tt 2000MHD} \\
\noalign{\smallskip}\hline\noalign{\smallskip}
cloud radius        & [pc]        & $R_0$                  & 0.99978 & 0.99978 & 0.99978 & 0.99978 \\
total cloud mass    & [$\msun$]   & $M_{\textrm{tot}}$     & 502.603 & 502.603 & 2152.11 & 2152.11 \\
mean mass density   & [g/cm$^3$]  & $\langle \rho \rangle$ & $4.256 \times 10^{-21}$ & $4.256 \times 10^{-21}$ & $1.822 \times 10^{-20}$ & $1.822 \times 10^{-20}$ \\
mean number density & [cm$^{-3}$] &$\langle n \rangle$     & 1188.98 & 1188.98 & 5091.14 & 5091.14\\
mean molecular weight &      & $\mu$ & 2.14 & 2.14 & 2.14 & 2.14 \\
temperature         & [K]    & $T$   & 10   & 10   & 10   & 10 \\
sound speed         & [km/s] & $c_{\textrm{s}}$ & 0.196 & 0.196 & 0.196 & 0.196\\
rms Mach number & & $\mathcal{M}$ & 6.01 & 6.01 & 6.01 & 6.01 \\
rms turbulent Alfvenic Mach Number & & $\mathcal{M}_A$ & 2.1 & 2.1 & 2.2 & 2.2 \\
mean freefall time & [Myr] & $t_{\textrm{ff}}$ & 0.74 & 0.74 & 0.370 & 0.370 \\
sound crossing time & [Myr] & $t_{\textrm{sc}}$ & 9.96 & 9.96 & 9.96 & 9.96 \\
turbulent crossing time & [Myr] & $t_{\textrm{tc}}$ & 1.66 & 1.66 & 1.66 & 1.66 \\
Jeans length & [pc] & $\lambda_{\textrm{J}}$ & 0.413 & 0.413 & 0.199 & 0.199 \\
Jeans volume & [pc$^3$] & $V_{\textrm{J}}$ & 0.294 & 0.294 & 0.033 & 0.033\\
Jeans mass   & [$\msun$] & $M_{\textrm{J}}$ & 4.42 & 4.42 & 2.13 & 2.13\\
magnetic field & [$\mu$G] & $B$  &  0 & 56.7 & 0 & 120.5\\
mass-to-flux ratio & & $\lambda$ &  $\infty$ & 1.17144 & $\infty$ & 2.35979 \\
rigid rotation angular frequency & [rad/s] & $\Omega_{\textrm{rot}}$ & 1.114e-14 & 1.114e-14 & 1.114e-14 & 1.114e-14 \\
rotational energy fraction & & $\beta_{\textrm{rot}}$ &  1.8 \% & 1.8 \% & 0.4\% & 0.4\% \\
\noalign{\smallskip}\hline\noalign{\smallskip}
\multicolumn{7}{c}{Numerical simulation parameters} \\
\noalign{\smallskip}\hline\noalign{\smallskip}
simulation box size & [pc] & $L_{\textrm{box}}$ & 1.99956 & 1.99956 & 1.99956 & 1.99956 \\
simulation box volume & [pc$^3$] & $V_{\textrm{box}}$ & 7.99471 & 7.99471 & 7.99471 & 7.99471 \\
smallest cell size & [AU] & $\Delta x$ & 50.3465 & 50.3465 & 50.3465 & 50.3465 \\
\noalign{\smallskip}\hline\noalign{\smallskip}
\multicolumn{7}{c}{Simulation outcomes} \\
\noalign{\smallskip}\hline\noalign{\smallskip}
final simulation time & [kyr] & $t_{\textrm{final}}$ & 179.3 & 232.2 & 42.8 & 49.6 \\
number of sink particles formed & & $n_{\textrm{sinks}}$ & 16 & 6 & 45 & 3\\
max sink mass & [$\msun$] & & 2.01528 & 9.53198 & 19.4442 & 31.2937 \\
min sink mass & [$\msun$] & & 0.0264274 & 0.164572 & 0.00759937 & 8.42947  \\
mean sink mass & [$\msun$] & & 0.696485 & 2.84511 & 0.680372 & 19.8616 \\
median sink mass & [$\msun$] & & 0.525414 & 1.56426 & 0.0548092 & 19.8616 \\
\hline \\
\end{tabular*}
\label{table:simulation_parameters}
\end{table*}

\subsection{Filament Identification}
\begin{figure*}[htbp]
\includegraphics[width=7.0in]{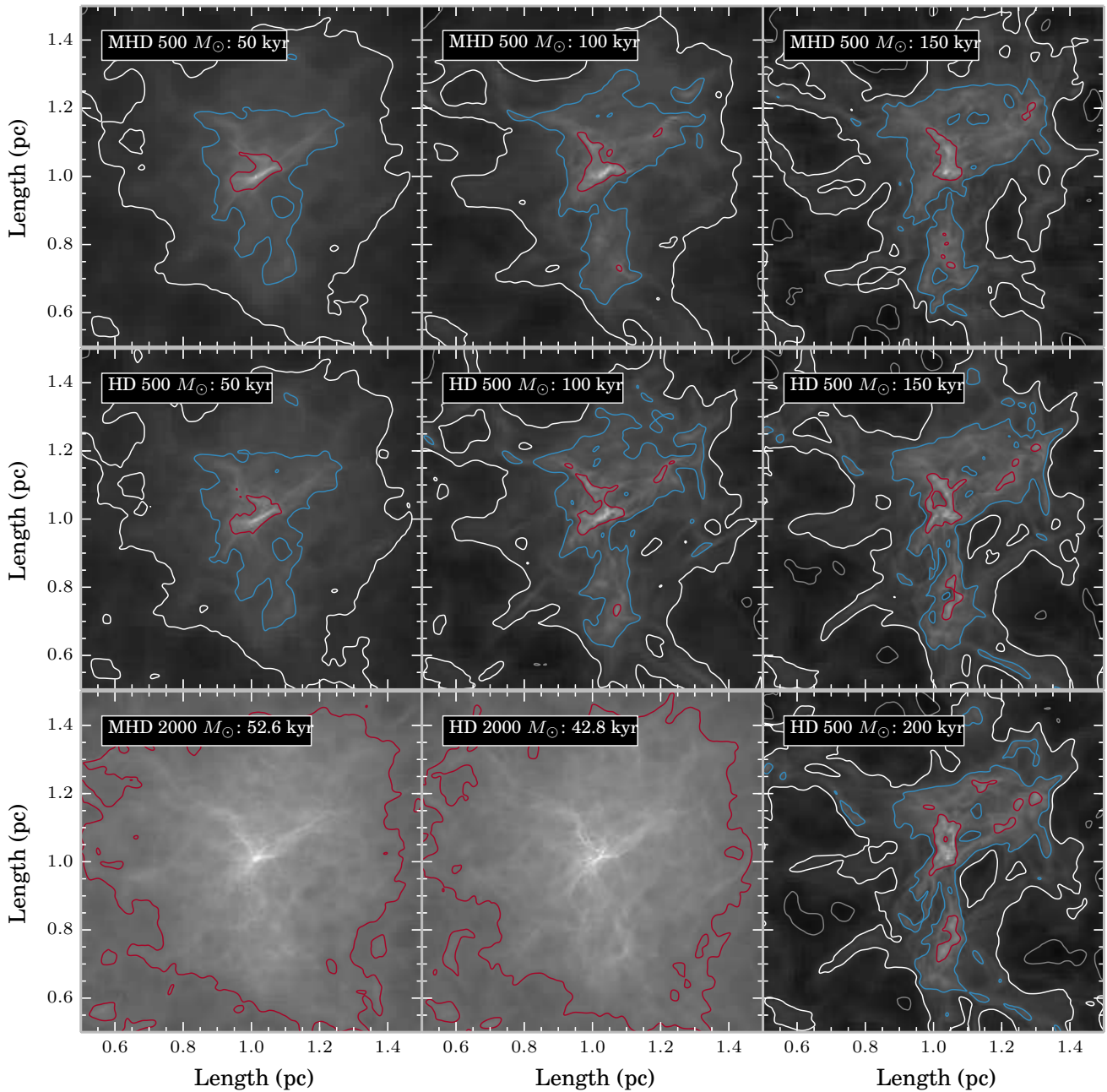}
\caption{A comparison of the column density distribution projected along
	the X axis for the simulations.  The top two rows show the 500~\Msol\ simulations
	at 0.05, 0.1, and 0.15~Myr after the start of the simulation (left to right) 
	for the MHD (top) and HD (middle) runs.  The bottom row shows the MHD and HD 
	2000~\Msol\ simulations at 0.05~Myr (left and middle) and the HD 500~\Msol\ 
	simulation at 0.2~Myr (right).  All simulations are shown cropped to the inner 
	1.5~pc to better show the smaller-scale structure that forms.  The greyscale range
	is the same in all panels, going from 0.01 to 10~g~cm$^{-2}$ 
	($\sim 3\times10^{21}$ to $3\times10^{23}$~cm$^{-2}$) from black to
	white, with a logarithmic scaling applied.  
	The overlaid contours show column densities of 0.02, 0.04, 0.075, and 0.2~g~cm$^{-2}$
	(5.3, 1.1, 2.1, and 53 $\times 10^{21}$~cm$^{-2}$)
	in grey, white, blue, and red, respectively.  {\it If} several assumptions 
	are made, including that all pixels belong to cylindrical structures
	with a characteristic width of 0.1~pc, then the contours also correspond to
	mass per unit length values 0.5, 1, 2, and 5 times the critical mass per
	unit length at a temperature of 10~K (18~\Msol~pc$^{-2}$).  
	Note that these assumptions are
	poor for regions not associated with filamentary structure.  
	See Section~3.4 for more detail.}
\label{fig_coldens}
\end{figure*}

The initial turbulent velocity field quickly results in a highly 
filamentary structure, as illustrated in Figure~\ref{fig_coldens}. 
We run each of our simulations until the 
filamentary structure is well-developed; the simulation is stopped at
0.2 to 0.3 free-fall times for the 500~\Msol\ simulations (for the 
MHD and HD simulations respectively), and 0.13 free-fall
times for the 2000~\Msol\ simulations. As \FLASH is an adaptive mesh 
refinement (AMR) code, we first take the output files and map them to 
a uniform grid, downsampling somewhat to allow the entire grid to fit 
into memory.  Even with the downsampling, our resolution is $\sim$0.002~pc,
much better than achievable with {\it Herschel} for nearby star-forming regions.
We then project the density along each of the coordinate 
axes to create column density maps.  

Figure~\ref{fig_coldens} shows
the column density in the X projection for both the 500~\Msol\ and 2000~\Msol\
simulations at all time steps analyzed.
Note that the MHD simulation was run for 0.15~Myr, while the HD simulation
was run for 0.2~Myr for the 500~\Msol\ simulations, giving one additional time step 
for our HD analyses.
In this figure, all the panels have the same dynamic range shown for the greyscale
column density, highlighting that material accumulates into filamentary structures
quite quickly (top and middle panel from left to right), and that having an initially higher
density more rapidly leads to dense filamentary structures due to the 
increased importance of gravity (bottom row, left and middle
panels).  Finally, the presence of a magnetic field acts to slow the accumulation of
material into dense filaments, as can be seen comparing the top and middle row panels,
or the bottom row left and middle panels.  We will return to this point in more
detail in Section~3 and beyond.

To extract the filamentary structure evident in Figure~\ref{fig_coldens} , 
we use the \disperse filament-finding 
algorithm\footnote{http://www2.iap.fr/users/sousbie/} described in 
\citet{Sousbie1} and \citet{Sousbie2}. 
The \disperse algorithm identifies persistent 
topological structures such as peaks, voids, and filaments, and is effective 
even if the image is noisy. It has been extensively used on {\it Herschel} 
observations for filamentary structure identification, e.g. 
\citet{Arzoumanian11,Schneider12,Peretto+2012,Palmeirim13}.
In {\sc DisPerSE}, there are several user-defined parameters to control
the resulting filamentary network: persistence and robustness thresholds, 
smoothing, and a maximum angle.  The two 
thresholds can be thought of as very roughly corresponding to
criteria for a minimum absolute brightness (persistence threshold) and
a minimum relative brightness compared to neighbouring features
(robustness threshold).  Smoothing removes small-scale `wiggles' from the
initial filament spine, while the angle is used to specify the
minimum angular rotation between two initial filament spine segments that can
be joined together and still be classified as the same filament.  Filament 
spine segments which meet at a right angle, for example, are likely not
part of the same filament.

We identify filaments using a persistence threshold of 0.025~g~cm$^{-2}$ 
and a robustness threshold of 0.05~g~cm$^{-2}$ in the 500~\Msol\ simulation
(or 7 and 14 $\times 10^{21}$cm$^{-2}$) and thresholds of 
0.1~g~cm$^{-2}$ and 0.2~g~cm$^{-2}$ (or 2.8 and 5.6$\times 10^{22}$~cm$^{-2}$)
respectively for the 2000~\Msol\ simulation, smoothing the resulting filaments 
1000 times, and allowing the initially identified 
filament segments to be connected for angles of less than 60 degrees
(relative to a straight line).
These parameters were chosen after
testing a range of values to determine which values produced a filamentary
structure that best matched visually-apparent structures.
All of these thresholds for \disperse\ are above the standard
`threshold for star formation' found in nearby molecular clouds of around
$\sim 5-7 \times 10^{21}$~cm$^{-2}$ \citep[e.g.,][]{Johnstone04,Konyves13}. 
Unlike the {\it Herschel} analyses, we applied \disperse
directly on the column density map.  Since our column density maps include
only the gas from the simulated star-forming clump, with no
potential contribution from other dense structures within the larger
cloud, we have less need than with {\it Herschel} data
to apply filament-enhancing algorithms.
Finally, we excluded several very short filaments that \disperse
initially identified - in order to accurately determine the 
filament profile (below), we set a minimum length of 0.1~pc.

\begin{figure*}[htb]
\includegraphics[width=7.0in]{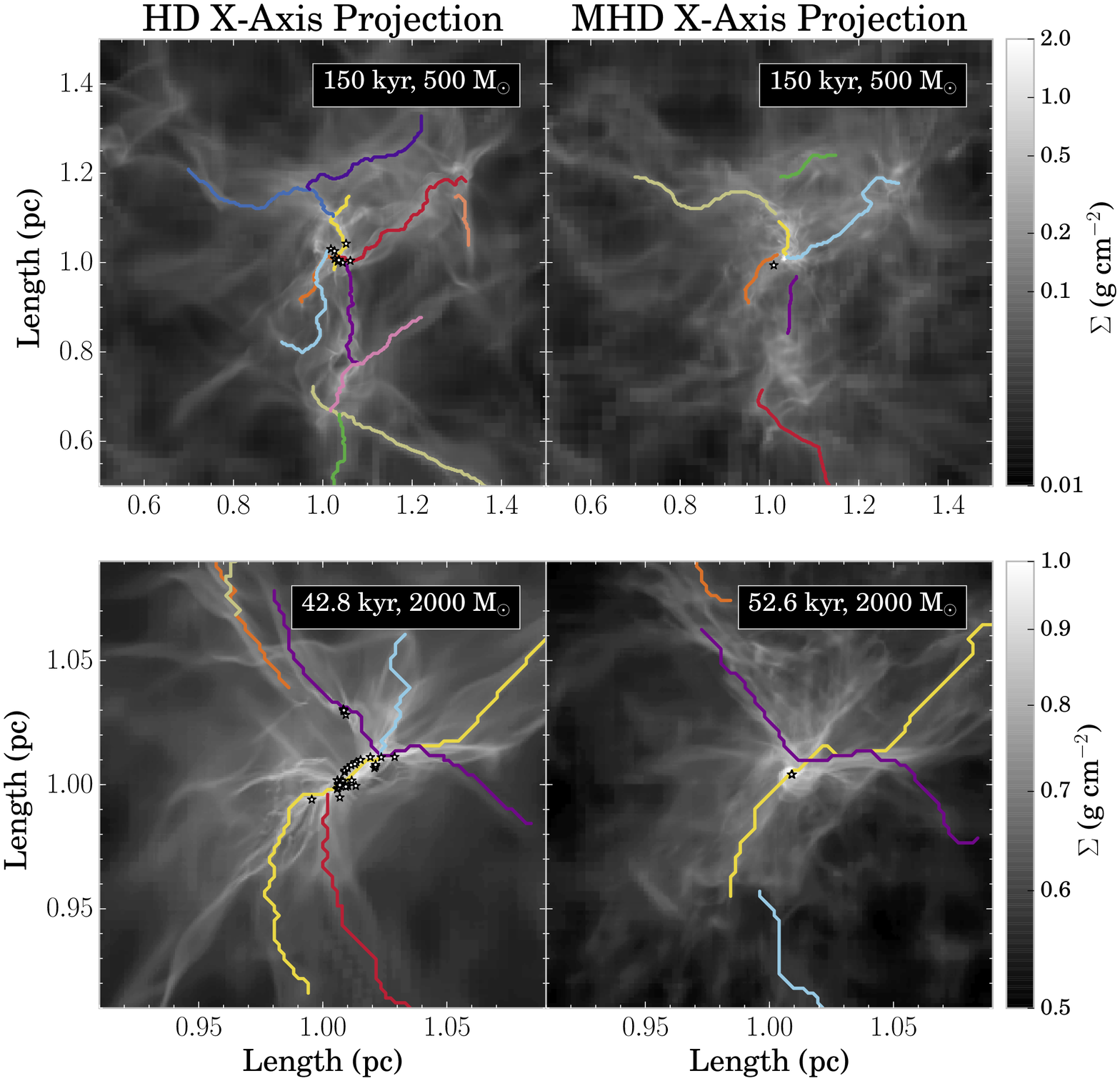}
\caption{ Examples of the filamentary structure identified in the simulations.
	The top panel shows the filaments identified using \disperse\ in the
	X projection of the 500~\Msol\ HD (left) and MHD (right) simulations
	at 0.15~Myr, while the bottom panel similarly shows the 
	2000~\Msol\ HD (left) and MHD (right) simulations at 0.05~Myr.
	Each coloured line indicates a unique filament spine identified
	in that projection.
	Note that the top and bottom panels zoom in to different extents
	to best illustrate the central filamentary structure; similarly,
	each row has a different greyscale scaling applied - see the scalebar on the right
	hand side.
	In all panels, sink particles formed at the specified time are shown
	by the white stars; in all cases, their formation is confined to
	the central clustered part of the simulation.
	}
\label{fig_fils}
\end{figure*}

Figure~\ref{fig_fils} shows the network of filaments identified 
in the X projection of the 500~\Msol\ and 2000~\Msol\ HD simulations 
overlaid on their column density maps.

One of the goals of our analysis is to track the time evolution of and
the effect of magnetic fields on individual filaments.  In order to do
so, \disperse was {\it not} used to identify a different network of
filaments at every time step and magnetic field value, as this
could potentially lead to different filaments being identified at 
different snapshots.  Instead,
for each of the three projections, we
started with the network of filaments identified with \disperse at
0.15~Myr in the HD simulation, and then searched for the corresponding
structures at different times and with magnetic fields present.
For the 2000~\Msol\ simulations, we instead started with the single
0.05~Myr time step.  We started with an automated procedure
to identify equivalent filaments at other time steps and / or with
magnetic fields, by searching for local column density maxima near
the reference set of filament spines.  After this step, 
all filament spines were verified and adjusted as
necessary by hand, using a combination of visual inspection of the
current column density snapshot and a movie of the time evolution of the
column density map for the 500~\Msol\ simulations.  
The simulations, particularly without the
moderating presence of magnetic fields, form significant substructure
on all scales, making it difficult to impossible for an automated
procedure to correctly `follow' the filaments in time and across
initial conditions.  

There are several cases where a filament could not be fully
traced to earlier times or in the corresponding simulation with
magnetic fields.  Some, but not all, of these cases appear to be
attributable to structures which are only apparent as filaments
in 2D due to a coincidence of independent 3D structures; at other time
steps, the real 3D structures have moved by different amounts
and no longer appear connected.  We include these structures
in our analysis where they do appear as a single filamentary
structure, as any real observation which only has column density information
is fallible to the same line of sight coincidence confusion.
We will address the full 3D nature of filaments in these simulations
in an upcoming paper.

A comparison of Figure~\ref{fig_coldens} and Figure~\ref{fig_fils} shows
that the filamentary network identified lies only in the very densest
part of the cloud, where the estimated mass per unit length value is
signficantly above the thermal critical value 
(white contours in Figure~\ref{fig_coldens}).  We will
return to this point further in Sections~3.2 and 3.3. 

\subsection{Calculating Radial Column Density Profiles}
\begin{figure}[htb]
\includegraphics[width=3.5in]{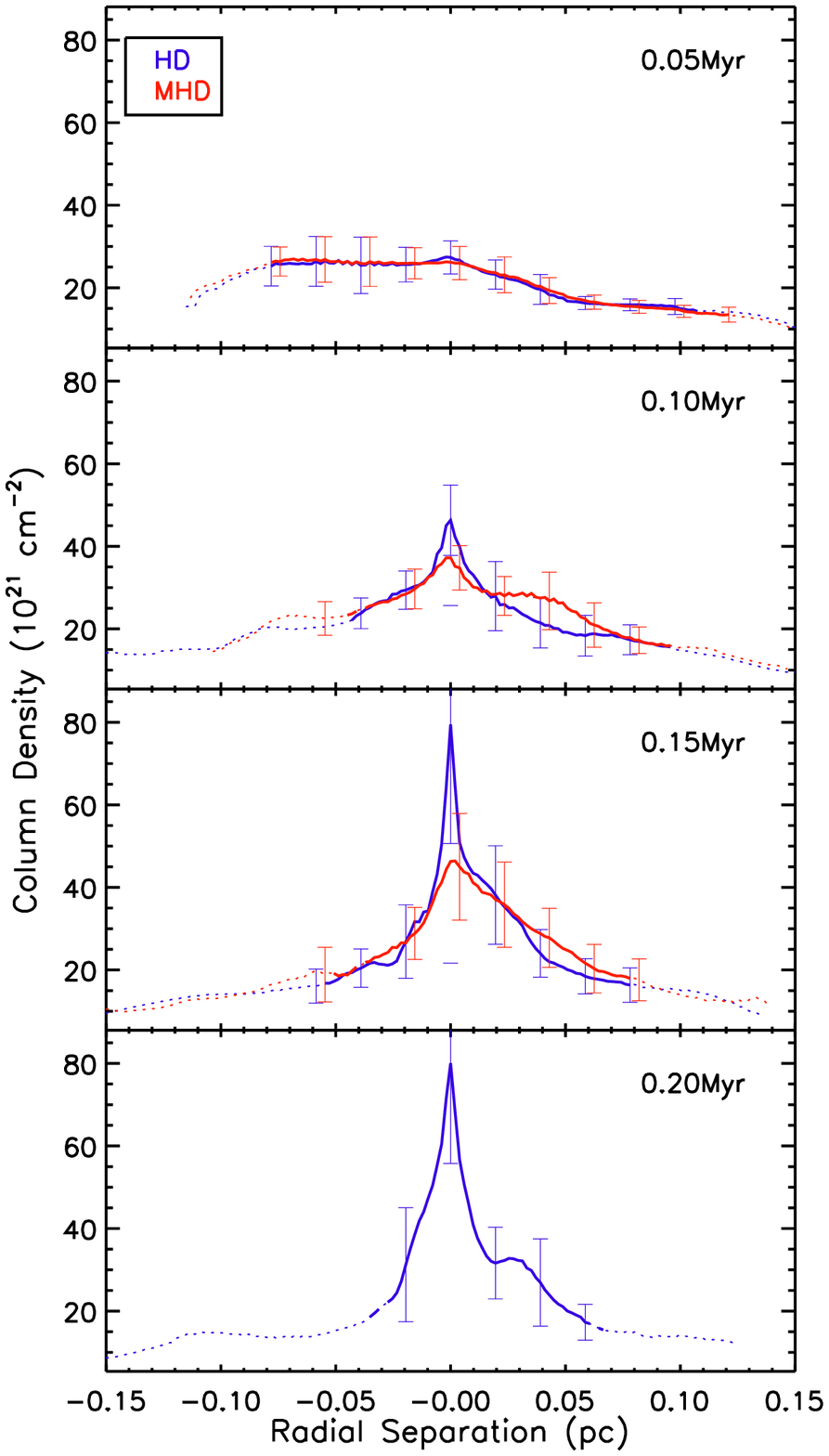}
\caption{Time evolution of the radial column density profile of a 
	filament identified in the Z projection of the 500~\Msol\ simulations.
	Solid lines show the reliable portion of the column density
	profile, while the dotted lines indicate less reliable measurements
	(i.e., data over less than 75\% of the length of the filament 
	was included).  The error bars indicate the standard deviation
	of column density values at that radial separation.  The MHD
	error bars are slightly shifted for better legibility.  In this
	example, the filament continues to contract in both the MHD and HD
	simulations, although at a faster rate in the HD case.}
\label{fig_radial1}
\end{figure}

\begin{figure}[htb]
\includegraphics[width=3.5in]{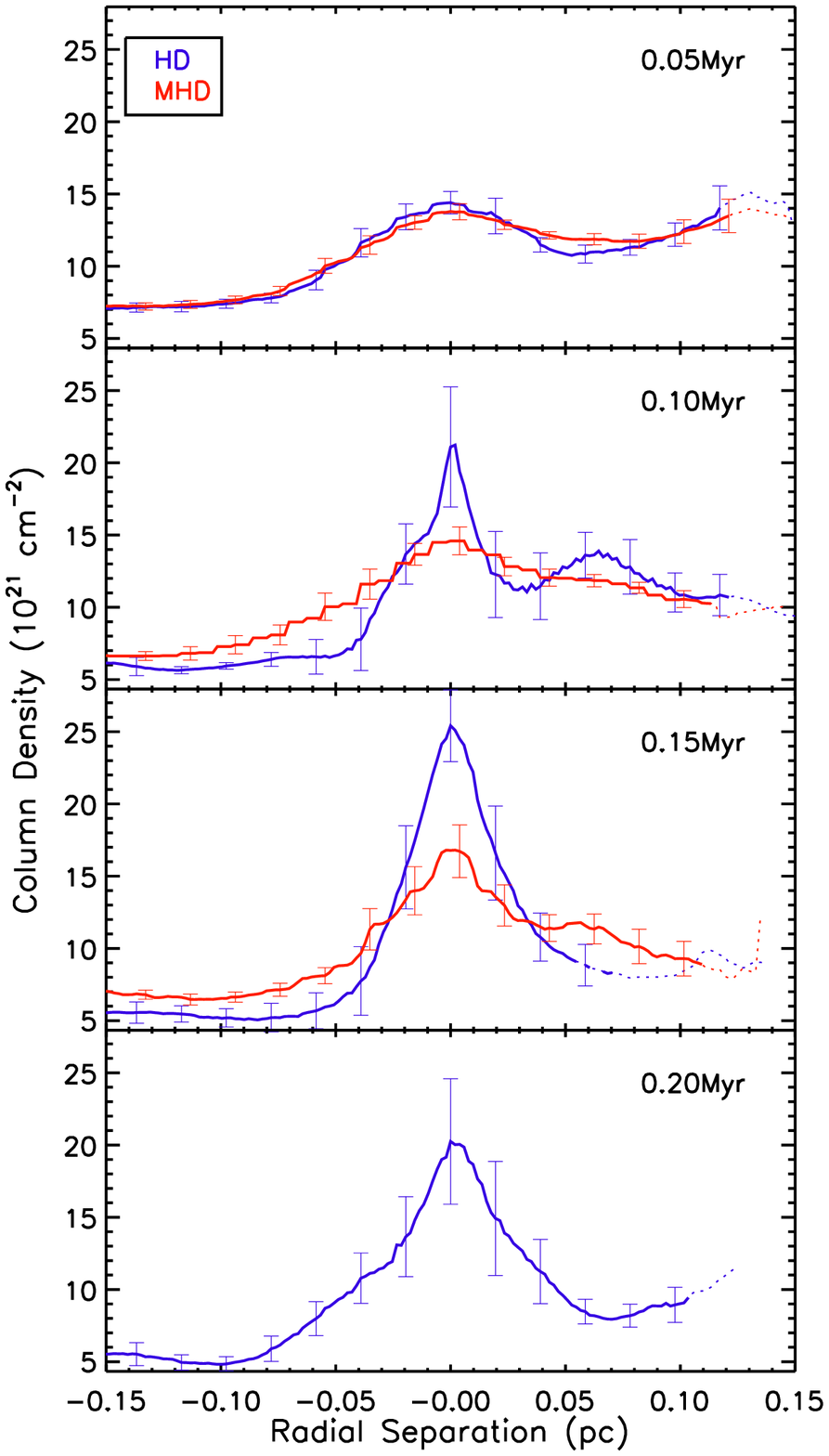}
\caption{Time evolution of the radial column denstiy profile identified
	in the Z projection of the 500~\Msol simulations.  See
	Figure~\ref{fig_radial1} for details on the plotting conventions 
	used.  In this example, the filament contracts and then expands
	in both the HD and MHD simulations.}
\label{fig_radial2}
\end{figure}

Once the filaments are identified, we measure the radial column density
profiles along them.  Since the filaments tend to converge toward the
simulation centre, and sometimes even intersect, care is needed to
properly calculate the radial column density profile.
First, we assign every pixel to the filament which it is closest to.
Next, we exclude pixels which lie very close ($< 0.01$~pc) to two
or more filaments -- this value was chosen to provide a 
balance between not including too many locations which might
provide non-representative measures of a given filament profile, and 
not excluding too large a fraction of material around the filaments.
We then calculate the mean column density of pixels in separation
bins equal to the pixel size ($\sim 0.002$~pc).  Finally, to ensure
that the filament profiles are accurate, we exclude
the measurement for any radial bin where at least 25\% of the total
length of the filament, at that separation, was not included in the
profile calculation.  This final criterion ensures that all 
radial column density profile measurements used in our analysis
are reliable - there are no cases where data from only a 
few pixels are used to infer the filament's properties.
We note that the above restrictions limit our analysis
to a smaller range in radii than used in \citet{Smith14}, although
the range is closer to \citet{Arzoumanian11}.  
\citet{Smith14} analyze only the brightest one or two filaments
in any given simulation snapshot, which ensures that the contamination
in filament profiles will be minimal; with our inclusion of 
fainter filaments, only smaller radial separations from a given
filament spine are free from material from neighbouring filaments.
Figures~\ref{fig_radial1} and \ref{fig_radial2} show several example radial column density 
profiles which will be further discussed in Section~4.


\section{Basic Filament Properties}
Visual inspection of the resulting radial column density profiles 
(e.g., Figures~\ref{fig_radial1} and \ref{fig_radial2}) reveals
a variety of characteristics.  
We expect that after the first turbulent shocks form
a filamentary structure, gravity acts to continue to concentrate mass
onto these filaments, leading to higher and narrower peaks with time.  
An initially higher mean density should 
increase gravity's pull and lead to a faster filament evolution.  
The presence of a magnetic field
should cushion the initial turbulent
compressions, reducing the amount of material initially in the filament,
and giving the appearance of `fluffier' filaments.
The subsequent evolution of MHD filaments should then be slowed relative to the
HD case by gravity's weaker pull on the the initial lower concentration 
of mass, and possibly also further action by the magnetic field, depending
on its orientation.

Broadly, these behaviours do hold -- the visual impression from
watching movies of each simulation suggest this behaviour,
nevertheless,
we find instances
of filaments dissipating over time, suggesting gravity was insufficient
to prevent the initial turbulent compression from re-expansion.
In some of these cases, magnetic fields appear to help to slow or
prevent this re-expansion, causing the HD filament to have a higher
and narrower peaked profile than in the MHD case.  In other instances,
data excluded for one or more of the reasons mentioned above (difficulty
in tracing the filament, or exclusion due to unreliability) also
prevents the full influence of time or magnetic fields to be fully
assessed.

Despite this more complex behaviour, there are still several simple
measures that we can make to gain insight into the behaviour observed.

\subsection{Filament Widths}
The conceptually simplest measureable filament property is its width.  Although filament
widths measured with {\it Herschel} span at least a factor of five
\citep[e.g., Figure~7 in][]{Arzoumanian11}, it is often stated that
filaments have a constant width of $\sim$0.1~pc.  
Note, however, that \citet{Juvela12b} find a larger scatter in filament
FWHM values in their analysis of (different) {\it Herschel} data, although
some of their filaments are much more
massive and / or more distant than the \citet{Arzoumanian11} sample.
We measure the width
of all of the filaments tracked in our simulations in the simplest
possible method -- the extent of the radial profile at half of the
peak value, i.e., the FWHM.  Table~\ref{tab_fwhm} shows our results,
separated by time step, magnetic field, and mass.  Included is the
mean and standard deviation of FWHM values measured, along with the
number of FWHM values considered.  Some filaments did not have reliable radial
column density profiles out to sufficiently large
radial separations to allow the FWHM to be measured; these were excluded
from the values given in Table~\ref{tab_fwhm}.
 
\begin{table*}
\tablenum{2}
\centering
\tabletypesize{\scriptsize}
\caption{Filament FWHM values \label{tab_fwhm}}
\begin{tabular}{cccccccccc}
Mass & 
Time &
\multicolumn{2}{c}{HD - FWHM stats$^{a}$} &
\multicolumn{2}{c}{MHD - FWHM stats$^{a}$} &
\multicolumn{2}{c}{HD - FWHM stats$^{b}$} &
\multicolumn{2}{c}{MHD - FWHM stats$^{b}$} \\
(\Msol) &
(Myr) &
mean(pc) &
stddev(pc) &
mean(pc) &
stddev(pc) &
mean(pc) &
stddev(pc) &
mean(pc) &
stddev(pc) \\
\tableline
 500 &  0.05 &  0.212 &  0.132 &   0.262 &  0.164  & 0.211 & 0.127 & 0.268 & 0.175 \\
 500 &  0.10 &  0.105 &  0.086 &   0.176 &  0.130  & 0.066 & 0.051 & 0.116 & 0.087 \\
 500 &  0.15 &  0.079 &  0.073 &   0.130 &  0.090  & 0.050 & 0.028 & 0.104 & 0.051 \\
 500 &  0.20 &  0.058 &  0.047 &     N/A &    N/A  & 0.044 & 0.025 &   N/A &   N/A \\
2000 &  0.05 &  0.119 &  0.113 &   0.168 &  0.147  & 0.132 & 0.137 & 0.170 & 0.142 \\
\tableline
\end{tabular}
\\
$^{a}$ All measureable filaments included.
$^{b}$ Only filaments where the FWHM could be measured at all times in HD and MHD were included.
	Note that the 500~\Msol\ and 2000~\Msol\ filaments are different samples.

\end{table*}

Although the dispersion is large, owing in part to the disparate behaviours
discussed earlier, it is clear that on average, the filaments do behave
as expected.  Filaments generally get narrower with time and in higher mean density
environments, and are wider when magnetic fields are present.  Furthermore,
this trend is somewhat subtle: all of the simulation snapshots give filaments
that have widths within the range that observers find.
\citet{Heitsch13a,Heitsch13b} note that in accreting filament models, the
filament width can remain relatively constant throughout much of a filament's evolution,
if either the ram pressure from accreting material is small \citet{Heitsch13a} or
in the case of a weakly magnetized accretion.  
\citet{Hennebelle13} propose
a model wherein ion-neutral friction dominating the dissipation of 
turbulence accounts for a relatively constant filament width of $\sim~0.1$~pc
while \citet{Gomez14} suggest a constant width may be caused by a 
balance between 
large-scale accretion onto filaments and accretion from the filaments
onto the dense cores and stars forming within them. 
Our simulations do not contain ambipolar diffusive effects for the
magnetic field, so that the substructure formed involves a balance between 
accretion, gravity, and turbulence.  Future work will measure the accretion 
onto filaments versus the dense cores.

We also tried fitting Gaussians to the filament profiles, with a constant
background level set as a free parameter (fits not shown), 
similar to other work \citep{Arzoumanian11,Smith14}\footnote{All 
of our fits (including those in Section~4) make use of the IDL 
mpfit routine by 
\citet{Markwardt09}, available at {\tt http://www.physics.wisc.edu/ \\
$\sim$craigm/idl/fitting.html}}.
Table~\ref{tab_fwhm_gauss}
shows the equivalent FWHMs derived from the Gaussian fits for the filaments.
Although the FWHM values tend to be smaller when using a Gaussian fit with a 
non-zero background, especially at earlier times (where the relative amplitude of 
the background is larger), the same general trends hold true: the filament width
decreases with time, and tends to be larger when magnetic fields are present.
The mean widths are still generally consistent with the observations,
although the range of widths is larger in the simulations, similar to
the findings of \citet{Smith14}.

\subsubsection{Biases}

\citet{Smith14} provide a detailed consideration of
factors which can impact the measured filament width.  For example, they
find that when performing a Gaussian fit, the best-fit FWHM strongly depends
on the radial extent of the profile: including measurements from larger
separations from the filament spine tends to increase the FWHM.  
Presumably this is at least partly caused by a lower background column
density being fit for profiles that extend further from the filament spine.
Our analysis tends to use a smaller radial extent than in \citet{Arzoumanian11}
and especially \citet{Smith14} due to potential contamination from other 
nearby filaments, which may explain why we tend to measure narrower
filament widths in our Gaussian fits\footnote{\citet{Smith14} focus their 
analysis on the brightest one or two filaments in each of their simulations,
which ensures that the contamination from other filamentary structures
will be more minimal, even with a larger radial extent.}.  
Measuring the
FWHM directly from the profile will be robust to radial range variations, but
has its own bias.  Unresolved central filament peaks become lower with poorer
resolution, which would change the peak column density used to estimate
the FWHM (R. Smith, priv. comm.).  
Both the biases in the FWHM and Gaussian-fitted width measurements are
primarily systematic, affecting absolute rather than relative values.
(We emphasize that this statement does not imply that the {\it range} 
in widths is invariant, but that the relative rankings likely are, i.e., the widest 
filaments appear to be the widest with any measure.)
We expect then that our conclusions about the weak trends in width are
therefore robust, a point supported by the similarity in behaviour using
either width measurement.

Finally, we note that the timescale over which we are able to analyze the 
filaments is relatively short: 0.2 to 0.3 times the global free-fall time
for the 500~\Msol\ simulations, and 0.13 times the global free-fall time
for the 2000~\Msol\ simulations.  Since the filaments form in the 
denser parts of the simulation, a larger number of {\it local} free-fall
times would have elapsed.  Nonetheless, analysis over a longer timescale
could reveal stronger signs of filament evolution than we are able to
probe here.

\begin{table*}
\tablenum{3}
\centering
\tabletypesize{\scriptsize}
\caption{Filament Gaussian-fit FWHM values \label{tab_fwhm_gauss}}
\begin{tabular}{cccccccccc}
Mass & 
Time &
\multicolumn{2}{c}{HD - FWHM stats$^{a}$} &
\multicolumn{2}{c}{MHD - FWHM stats$^{a}$} &
\multicolumn{2}{c}{HD - FWHM stats$^{b}$} &
\multicolumn{2}{c}{MHD - FWHM stats$^{b}$} \\
(\Msol) &
(Myr) &
mean(pc) &
stddev(pc) &
mean(pc) &
stddev(pc) &
mean(pc) &
stddev(pc) &
mean(pc) &
stddev(pc) \\
\tableline
500  & 0.05 & 0.08 & 0.07 & 0.10 & 0.07   & 0.11 & 0.10 & 0.12 & 0.09 \\
500  & 0.10 & 0.05 & 0.03 & 0.09 & 0.05   & 0.06 & 0.03 & 0.12 & 0.06 \\
500  & 0.15 & 0.04 & 0.02 & 0.06 & 0.03   & 0.05 & 0.02 & 0.07 & 0.03 \\
500  & 0.20 & 0.05 & 0.02 & N/A  & N/A    & 0.05 & 0.02 & N/A  & N/A  \\
2000 & 0.05 & 0.04 & 0.03 & 0.10 & 0.11   & 0.05 & 0.03 & 0.15 & 0.15 \\
\tableline
\end{tabular}
\\
$^{a}$ All measureable filaments included.
$^{b}$ Only filaments where a Gaussian could be fit at all times in HD and MHD were included.
	Note that the 500~\Msol\ and 2000~\Msol\ filaments are different samples.

\end{table*}

\subsection{Mass per Unit Length}
\label{sec_ML}

The simplest equilibrium model for a filament is that of the isothermal cylinder,
presented in \citet{Ostriker64}, where gravity is balanced by thermal pressure along an 
infinite cylinder.  In this model, the stability
of the cylinder is controlled by the mass per unit length, $M_{line}$.  
The critical mass per unit length, in turn, depends only on 
the temperature:
\begin{equation}
M_{\textrm{line}}^{\textrm{crit}} = \frac{2 k_B T}{\mu m_H G} \equiv \frac{2 c_s^2}{G},
\end{equation}
where $c_s$ is the sound speed and $G$ the gravitational constant
\citep{Ostriker64}.
Furthermore, \citet{InutsukaMiyama1997} showed that isothermal filaments are 
unstable to axisymmetric perturbations of wavelength greater than about 2 
times the filament diameter if the mass per unit length is close to this 
critical value.

In our simulations, the temperature is
set at a constant 10~K, which implies $M^{crit}_{line} = 18$~\Msol~pc$^{-1}$.
Turbulent motions can also provide additional support through 
raising the typical velocity dispersion of the gas above the
thermal value;
\citet{Heitsch13a} points out that non-thermal motions
can be driven by the accretion of material onto the
filament itself \citep[see also][]{Peretto14}.
\citet{FiegePudritz00a} show that a more appropriate
critical mass per unit length value is given by
\begin{equation}
M_{\textrm{line}}^{\textrm{crit}} = \frac{2 \langle \sigma^2 \rangle}{G},
\end{equation}
where $\langle \sigma^2 \rangle$ is the velocity dispersion including both 
thermal and non-thermal components. 
A careful analysis of the velocity fields would be required to 
determine precisely how much nonthermal support is provided on
the scales of interest; the approximation often assumed is
\begin{equation}
\sigma^2 = c_s^2 \times (1 + \mathcal{M}^2/3)
\end{equation}
\citep[e.g.,][]{Klessen00}.  With $\mathcal{M} \sim 6$ in our simulations,
that would lead to raising \Mcrit\ by a factor of roughly 37, 
giving 670~\Msol~pc$^{-1}$.  

In the case of a magnetized turbulent cloud, there is a magnetic correction 
that must be made to eqn~3.  In the case that there is only a poloidal magnetic 
field and no toroidal (wrapped field) component, the magnetic pressure helps support 
the filament against gravity.  Taking eqn 27 of \citet{FiegePudritz00a} with 
$\Gamma_{\phi} = 0$ (no toroidal field), and using \citet{FiegePudritz00a} eqn~23
to convert between the vertical magnetic field flux and magnetic field strength, the
critical mass per unit length becomes 
\begin{equation} 
M_{line}^{(crit,B)} = M_{line}^{crit} \times (1 + \mathcal{M}_A^2/2)  
\end{equation} where $\mathcal{M}_A = v_A/\sigma$ is the turbulent Alfven Mach number whose 
Alfven speed, $v_A$, is $B/\sqrt{4\pi \rho}$.
For super-Alfvenic turbulence, $\mathcal{M}_A < 1$, this correction is slight.   In our 
simulation, however, 
$\mathcal{M}_A$ is $2.1 - 2.2$ (see Table~1).
Thus, our MHD turbulence is somewhat sub-Alfvenic, i.e., the magnetic field strength dominates
the turbulence, and this implies that the 
critical line mass for our models is somewhat greater than the purely hydrodynamic 
turbulent case; $M_{line}^{(crit,B)} \sim 3 M_{line}^{crit}$.  
This result predicts that our MHD case should be considerably less susceptible 
to fragmentation than the hydro case.  We note that as the turbulence is damped 
and the line width reduces to the thermal value, the relative magnetic contribution to the 
(thermal) line mass can become significantly more important depending on the orientation 
of the field across the filament.  

There has been little time in these
simulations for the turbulence to decay, but some of the power of 
the turbulence
is on larger scales than our filaments, so the turbulent critical mass
per unit length of 670~\Msol~pc$^{-1}$ is an 
upper limit to
the true effective critical mass per unit length of the simulated filaments.
Indeed, the observed velocity dispersion in filamentary gas tends to be
only of order twice the sound speed in nearby filaments that have peak column
densities similar to those formed in our simulations 
\citep[e.g.,][]{Hacar11,Hacar13,Arzoumanian13,Kirk13}. \citet{Arzoumanian13} furthermore
find that for filaments with masses per unit length much higher than the critical thermal
mass per unit length, the velocity dispersion increases with increasing mass per unit
length, a trait they attribute to infall of material onto the filaments.
In our simulations, most if not all of of the turbulent motion is likely due to the
remnants of the initial turbulence, given there has been little time for that to decay,
or for infall to generate additional turbulent motions.
Assuming the total filament velocity dispersion is twice the thermal value would 
increase \Mcrit\ by a factor of four.
The estimated mass per unit length contours shown on Figure~\ref{fig_coldens} illustrate that
most of the dense filamentary structure is found in areas with $M^{crit}_{line}$ above 4-5
for the 500~\Msol\ simulations, and even higher in the 2000~\Msol\ simulations.
\citet{Peretto14} estimated that non-thermal support in
SDC13 would contribute to 
lowering the mass per unit length values in the filaments
to 1--2 times the critical value from 4--8 times the critical value
if non-thermal motions were not accounted for.
Note that depending on the orientation, magnetic fields could either
aid or hinder gravitational collapse.

The degree of gravitational fragmentation of the filaments can, to some degree, 
be ascertained by the number of sink particles that form in the simulations.  It is 
notable that the sinks are typically found in or near the filaments.  It is also clear 
that the number of sinks that form in MHD simulations is smaller, sometimes notably so, 
in comparison with their hydrodynamic counterparts.  Thus, using the data in Table~1, 
we see that in the 500~\Msol\ simulation, while 16 sinks particles appear in the HD run, 
only 6 are apparent in the MHD case.  The suppression is even greater in the 2000~\Msol\
simulation (although in that case the runtime was shorter) where 45 formed in the HD 
case as compared to 3 in the MHD case.  Clearly, magnetic support is significantly reducing
fragmentation.

Non-isothermal equilibrium cylinder models have also been investigated
\citep[see][and the discussion therein]{Recchi13}.  In the case of a
thermally-supported equilibrium cylinder, where the temperature gradually
increases outward, similar to observations, \citet{Recchi13} find that
the mass per unit length which can be supported is only about 20 to 30\%
larger than in the isothermal case.  Since the simulations we analyze are
strictly isothermal, we do not consider this class of models further here.

\subsection{Individual filament M/L measurements}
Measuring the total mass of each filament is difficult
as can be seen in Figures~\ref{fig_coldens} and \ref{fig_fils}, 
the filaments
do not tend to have clearly defined outer boundaries.  This challenge is
exacerbated by the fact that many of the filaments lie close to one another --
the total mass cannot be derived by including material arbitrarily far away from the
filament's spine.  We determined that the best way
to estimate each filament's mass was to include only material within the FWHM
of the filament spine.  While this will necessarily provide a lower limit to
the true filament mass, lower and wider thresholds, such as the full width at
quarter maximum, cannot be determined for too large a fraction of the total
filament population (see discussion above).  Using a constant width for the
mass determination would bias the estimates toward relatively lower values
for the wider / fluffier filaments.  We note that instead adopting a filament
width based on the Gaussian fits discussed in Section~3 yields a similar behaviour
to that discussed below.

Using the estimated filament mass per unit lengths, we can test whether
this metric is a useful predictor of filament stability.  A very simple
proxy for the evolutionary path of a filament is to compare the peak
column density (in the radial column density profile) at two neighbouring
time steps.  If the filament is contracting or accreting, the second peak should be 
higher than the first, while the reverse would be true for an expanding
filament.\footnote{We verified this assumption by comparing the ratio of
filament FWHM values at subsequent times and found very good correlation: over
86\% of filaments interpreted as contracting or expanding based on their peak
column density ratio at neighbouring time steps show the same signature in their 
FWHM ratio.  Allowing
for slight measurement uncertainties (5\% error in the ratios) gives an
agreement between the two ratio measurements of just over 95\%.}  
We would therefore expect that higher mass per unit length
values (above the critical value) would correspond to a ratio in
peak column densities of greater than one (for the later time 
divided by the earlier time).  

\begin{figure}[h]
\includegraphics[width=3.5in]{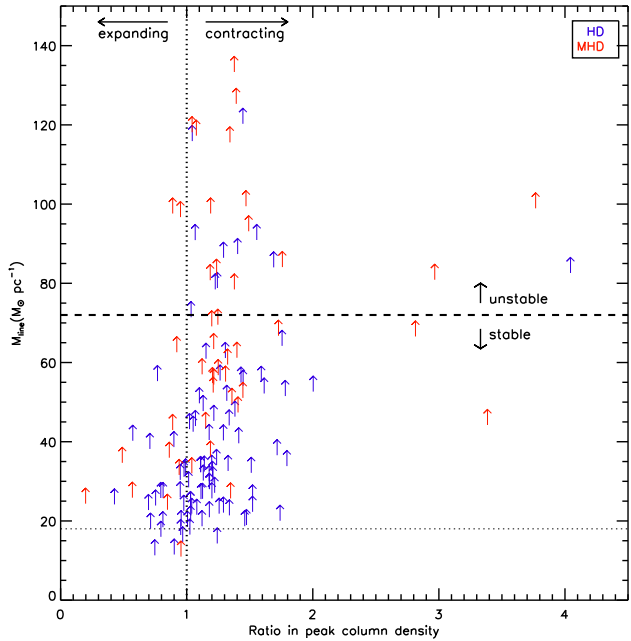}
\caption{The mass per unit length measured for a filament at a given time step versus
	the ratio in peak column densities for the subsequent versus current time step.
	Filaments which 
	are contracting would be expected to have a ratio in peaks
	greater than one (right of the vertical dotted line), whereas filaments 
	which are re-expanding would have a ratio in peaks of less than one (left of the
	vertical line).  The thick horizontal dashed line indicates the estimated 
	effective mass
	per unit length (assuming equal thermal and non-thermal contributions); filaments
	with values above this line would be expected to be contracting due to
	gravity, whereas those with lower values are stable against gravitational collapse.
	The thin horizontal dotted line indicates the critical mass per unit
	length assuming just thermal support; nearly all of the filaments have estimated
	mass per unit length values well in excess of the thermal critical value.  Note
	that the estimated mass per unit length values are all lower limits
	(see text for details).
	}
\label{fig_ml_vs_pks}
\end{figure}

Figure~\ref{fig_ml_vs_pks} shows 
the mass per unit length of each filament compared to the ratio of the
peak column density at the subsequent and current time steps.  The thermal
critical mass per unit length for a temperature of 10~$K$ is $\sim 18 $\Msol\ pc$^{-1}$, 
below nearly all of our measurements.  
Non-thermal motions likely contribute some amount
of support (Section~3.2).  In Figure~\ref{fig_ml_vs_pks} we show the critical
mass per unit length assuming the total velocity dispersion is twice the
thermal value, which gives a value of $\sim$72~\Msol\ pc$^{-1}$.  While
this is a rough approximation, it 
appears to denote the level above which no filaments are
found to be expanding.  Regardless of precisely where the effective critical mass
per unit length is drawn, we note a surprising result: many points occupy the
bottom right quadrant, contracting filaments whose
current mass per unit length implies 
gravitational stability.  Similarly, there are several filaments
whose mass per unit length ratio suggests graviational instability which 
are instead expanding.  Although some of
the presently contracting, low mass per unit length filaments (bottom right)
may expand at time steps beyond what we can trace, the figure highlights
the fact that predictions about the future evolution of filaments are
incomplete when only column density information is available.
This is also apparent in Figure~\ref{fig_radialfits_bkgrd}, where
the filament with the initially higher peak column density is the one
which later re-expands.
Although the individual mass per unit length values are not a good
predictor of future evolution, the fact that there is a weak correlation
between mass per unit length and peak column density ratio suggests
that some (limited) insight into the bulk behaviour of filaments 
can be gained from the simple isothermal-with-turbulence model.

\subsection{Region-wide Mass per unit Length}
Finally, we can also get a rough idea
of the stability of {\it all} the material in the simulation.
The {\it Herschel} Gould Belt team has provided an estimate
of the mass per unit length at every pixel, for material in their
curvelet map, i.e., that associated with structures having long axis ratios
\citep[e.g.,][]{Andre10}.
These mass per unit length estimates are made under the assumption that
each pixel is part of a filament or cylindrical structure with a typical width of
0.1~pc.  
The contours in Figure~\ref{fig_coldens} can be interpreted under a similar
set of assumptions, although we emphasize that since our calculation includes the entire
mass in the simulation, the equivalent mass per unit length values should not be
over-interpreted in regions not associated with filamentary structure.
The thermal critical value assumed for the contours in Figure~\ref{fig_coldens}
is that for a 10~K medium, i.e., \Mcrit $ = 18$~\Msol~pc$^{-1}$.
Most of the mass in filaments in the simulation lies above the critical mass per unit
length value; if thermal pressure was the only source of
support preventing gravitational collapse, we would expect to find
sink particles forming throughout the simulation.  Instead, all
of the sink particles form at mass per unit length values of 
at least 5 (red contour in Figure~\ref{fig_coldens}).
Including nonthermal motions, as discussed earlier, raises the critical
mass per unit length, therefore decreasing the ratio of the mass
per unit length to the critical value.  The mass per unit length
of individual filaments above which only contracting filaments are found, as
discussed in Section~3.3, was a similar factor (4) above the thermal critical value.
It therefore appears that only the densest parts of the simulation,
where filaments are present, are likely to be sufficiently dense 
for gravitational collapse to occur.

\section{Model Comparisons}
We next compare the radial column density profiles obtained to several cylindrical
equilibrium models: the isothermal cylinder, modified isothermal
cylinder, and pressure confined isothermal cylinder, described in detail
below.  In the simplest model, the isothermal
cylinder, thermal pressure balances gravity along an
infinite cylinder, leading to a 3D density profile decreasing as $r^{-4}$ at large radii,
or the column density varying as $r^{-3}$ \citep{Ostriker64}.
{\it Herschel} teams have found that the observed column density
profiles are shallower than the isothermal cylinder model and 
and instead use a modified profile, also referred to as a `Plummer-like' profile
\citep{Nutter08,Smith14,Plummer11}, 
where the power law exponent is an additional fitted parameter:
\begin{equation}
\Sigma(r) = A_p \frac{\rho_c R_{fl}}{\big(1 + (r/R_{fl})^2\big)^{\frac{p-1}{2}}}
\end{equation}
Here, $\Sigma$ is the column density, $\rho_c$ is the central density, 
$R_{fl}$ represents the scale of the inner flat portion of the profile,
$p$ is the power law index (with a value of 4 for the original Ostriker
model), and $A_p$ is a geometrical factor given by:
\begin{equation}
A_p = \frac{1}{cos i} \int^{\infty}_{-\infty} \frac{du}{(1+u^2)^{p/2}}
\end{equation}
where $i$ is the (unknown) inclination of the filament on the plane of the sky,
assumed to be 0 \citep{Arzoumanian11}.
The best-fitting value of $p$ often tends to range between 1.5 and 2.5, a
much shallower drop-off than in the $p=4$ \citet{Ostriker64} model
\citep[e.g.,][]
{Alves98,Lada99,Arzoumanian11,Malinen12,Juvela12,Hill12,Palmeirim13}.  
Some filaments, however, have been observed with 
column density profiles which are consistent with a $p=4$ isothermal model, 
\citep[e.g.][]{Nutter08,Pineda11,Hacar11,Bourke12}, while \citet{Contreras13} find
$p=4$ provides a good fit around star-forming clumps and $p=2$ is better
in the inter-clump areas.
Theoretically, shallow radial column density profiles are consistent with
equilibrium isothermal cylinder models that include helical magnetic fields
\citep{FiegePudritz00a}.  \citet{Smith14} show that $p\sim2$ profiles are
the norm for prominent filaments in hydrodynamic simulations without
magnetic fields, regardless of the type of turbulence considered.

We also applied the equilibrium model of \citet{Fischera12}, 
in which pressure from the medium surrounding the filament is also
included in the force balance.  In this analytic formulation,
the two quantities of interest are $P$, the pressure from the
external medium, and $f$, the ratio of the mass per unit length
to the critically stable value for the Ostriker cylinder.  
The full profile is given by:
\begin{align}
N_H(x) = \sqrt{\frac{P}{4\pi G}} \frac{\sqrt{8}}{1-f} (\mu m_H)^{-1}
	\times \frac{1-f}{1-f+x^2f}
	\notag \\
	\Bigg( \sqrt{f(1-f)(1-x^2)} +
	\sqrt{\frac{1-f}{1-f(1-x^2)}}
	\notag \\
	\times arctan\sqrt{\frac{f(1-x^2)}{1-f(1-x^2)}} \Bigg)
\end{align}
where $N_H$ is the column density in number units, $x$ is a 
scaled radial coordinate, $G$ is the 
gravitational constant, $\mu$ is the mean molecular weight, and $m_H$ is
the mass of a hydrogen atom \citep{Fischera12}.
The main effect of the pressure, $P$ is on the height of the central 
column density peak, while $f$ controls the shape of the
profile \citep{Fischera12}.  [The temperature of the gas is also
fit as part of the scale factor converting the radial coordinate $x$ into a physical radial
separation.]  In their re-analysis of the {\it Herschel}
filaments in Polaris, IC~5146, and Aquila, \citet{Fischera12}
find that their pressure equilibrium model also provides a good fit
to the filaments, 
with external pressures consistent with the range expected in the
ISM.

In all of the models, a non-zero background
column density can be included as a free parameter, i.e., 
\begin{equation}
\Sigma(r) = \Sigma_{model}(r) + \Sigma_0
\end{equation}
where $\Sigma_0$ is a constant.

Most if not all of the filament analyses include a background column density
term \citep[e.g.,][]{Arzoumanian11,Juvela12,Fischera12,Palmeirim13,Smith14}; 
{\it Herschel} analyses in fact allow the background to be fit by 
a linearly varying background column density: $\Sigma(r) = \Sigma_{model}(r) + \Sigma_0 +\Sigma_1(r)$
\citep[see Appendix B of][]{Palmeirim13}, although the fits are generally similar
when just a constant background column density is adopted (D. Arzoumanian, priv. comm.). 
We tried fitting the profiles with and
without a background column density and found that including the background
generally produced superior fits.  In the case of the modified isothermal
cylinder model (eqn 5), including a background term decreased the 
central density,
and allowed for a narrower peak to be fit, better matching the 
filament profiles.  The difference was most pronounced for the case of
the pure isothermal cylinder model (eqn 5 with $p=4$), 
where few cases produced good fits 
without a background column density.  The pressure confined 
model (eqn 7) almost never
converged to a satisfactory fit without a background column density term.
Appendix~A shows the results of fits with no background column density included.

\begin{figure}[htb]
\includegraphics[width=3.5in]{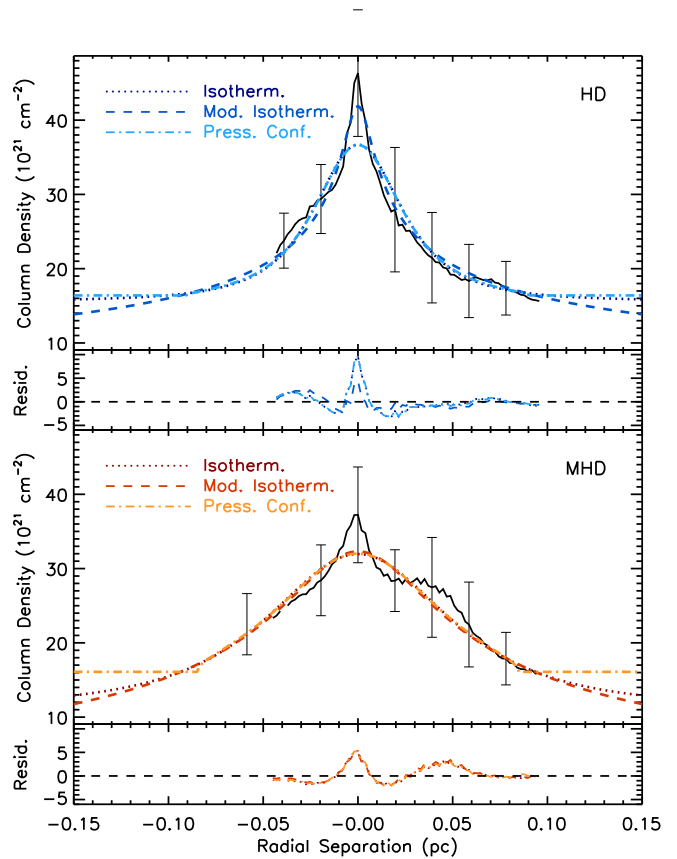}
\caption{The radial column density profile for the filament in Figure~\ref{fig_radial1} at
	0.1~Myr, including the best-fit models with
	a background column density term. 
        The top panel
        shows the profile and fits for the HD simulation, as well as the residuals
	between the models and simulation, while the bottom panel shows
        the same for the MHD simulation.  The solid black line indicates the mean column
        density at each radial separation, while the error bars denote the standard
        deviation in values at various radial separations.  The model lines shown are
        the purely isothermal cylinder (darkest, dotted line), the modified isothermal
        cylinder (less dark, dashed line), and the pressure confined cylinder
        (lightest, dash-dotted line).}
\label{fig_radialfits_bkgrd}
\end{figure}

Figure~\ref{fig_radialfits_bkgrd} shows the best fit models for one example filament at
0.1~Myr.
Following the general behaviour seen in the simulations,
the filament in the MHD simulation is `fluffier' (wider and lower
peak column density) than in the HD simulation.
This is a consequence of the significant amount of magnetic pressure support of the 
filaments as noted earlier.

\subsection{Modified Isothermal Cylinder Model Fits}

As can be seen in Figure~\ref{fig_radialfits_bkgrd}, the 
isothermal and modified isothermal cylinder model provide
near-identical fits, when a background column density
term is included.  In both the HD and MHD profiles shown,
the models have peak values within the errors of the 
simulated radial profile.  
As shown in Appendix~A, the models differ by
a greater amount when the background column density is fixed to
zero, with the pure isothermal model then generally providing a very
poor fit.
 
Table~\ref{tab_arz_bkgrd} gives the median model fit parameters
for filaments where a fit was possible at every time step in both the HD and MHD
simulations, to allow for any trends in the time evolution to be followed.
The typical powerlaw slopes found ($p \sim 1.3-2$) are similar to the best-fit
values in \citet{Arzoumanian11}. 
The typical dispersion (standard deviation) in the model
fit parameters is of the same size as the median values given in the table;
in all cases, there is no clear trend in the best fit parameters evolving as a function
of time.  
\citet{Juvela12b} and \citet{Smith14} point out that the 
modified isothermal fit is partially degenerate between the fit parameters,
which could hide evolutionary trends.  \citet{Smith14} demonstrated that
while no time evolution was seen in their best-fit parameters when all were free
to vary, fixing $R_{fl}$ gave best-fit central densities which clearly increased
with time.  We performed the same test and found a similar result, as shown in
Table~\ref{tab_arz_bkgrd_fix}, although the combination of expanding and contracting
filaments at later time steps diminishes the strength of the evolutionary trends.

\begin{table}
\tablenum{4}
\centering
\tabletypesize{\scriptsize}
\caption{Best fit parameters for modified isothermal cylinder, with background \label{tab_arz_bkgrd}}
\begin{tabular}{cccccccccc}
Mass & 
Time &
\multicolumn{4}{c}{HD $^{a}$} &
\multicolumn{4}{c}{MHD $^{a}$} \\
(\Msol) &
(Myr) &
$\rho_c$ &
$R_{fl}$ &
$p$ &
$\Sigma_0$ & 
$\rho_c$ &
$R_{fl}$ &
$p$ &
$\Sigma_0$ \\
\tableline
 500 &  0.05 &  0.9  & 3.0 & 2.3 & 5.4  &  1.4 & 3.0 & 2.4 & 7.7 \\
 500 &  0.10 &  14.3 & 0.6 & 1.5 & 3.2  &  5.5 & 2.0 & 1.8 & 3.5 \\
 500 &  0.15 &  8.8  & 0.8 & 1.8 & 8.7  &  1.7 & 3.6 & 3.1 & 4.5 \\
 500 &  0.20 &  4.5  & 2.3 & 4.2 & 9.7  &  N/A & N/A & N/A & N/A \\
2000 &  0.05 &  8.4  & 1.0 & 1.8 &  29  &  6.7 & 1.1 & 1.4 & 23 \\
 500 & all$^b$ & 4.6 & 1.7 & 2.0 &  4.6 & 1.7  & 2.7 & 2.0 & 4.3 \\
2000 & all$^b$ & 8.4 & 1.0 & 1.8 &  29  & 6.7  & 1.2 & 1.4 & 23\\
\tableline
\end{tabular}
$^{a}$~Mean of the best fit values for filaments fit at all times, with a non-zero
	background allowed for the fit: the
	central density, $\rho_c$ (in units of $10^5$~cm$^{-3}$),
	the central flat radius, $R_{fl}$ (in units of 0.01~pc), 
	the exponent, $p$, and $\Sigma_0$, the background column density term 
	(in units of 10$^{21}$~cm$^{-2}$); the standard deviation is often comparable
	in magnitude to the mean, with values of 0.6--18, 0.4--2.4, and 
	0.2-3.5 for $\rho_c$, $R_{fl}$ and $p$ respectively, in the same units.
	The background column density tends to have standard deviations larger
	than the mean value, and the mean is usually significantly larger than
	the median. \\
$^{b}$~Values of all profiles where a fit was possible are included here,
        i.e., relaxing the requirement of a fit for both HD and MHD,
        and for the 500~\Msol\ simulations, a fit at all time steps.
\end{table}

\begin{table}
\tablenum{5}
\centering
\tabletypesize{\scriptsize}
\caption{Best fit parameters for modified isothermal cylinder, with background, $R_{fl}$ fixed
at 0.01~pc. \label{tab_arz_bkgrd_fix}}
\begin{tabular}{cccccccc}
Mass & 
Time &
\multicolumn{3}{c}{HD $^{a}$} &
\multicolumn{3}{c}{MHD $^{a}$} \\
(\Msol) &
(Myr) &
$\rho_c$ &
$p$ &
$\Sigma_0$ & 
$\rho_c$ &
$p$ &
$\Sigma_0$ \\
\tableline
 500 &  0.05 &  1.6  & 1.6 & 3.2  &  1.6  & 1.4 & 2.5 \\
 500 &  0.10 &  2.7  & 1.8 & 7.0  &  1.9  & 1.5 & 2.6 \\
 500 &  0.15 &  3.3  & 2.1 & 8.6  &  2.2  & 1.6 & 4.2 \\
 500 &  0.20 &  3.6  & 2.0 & 7.3  &  N/A  & N/A & N/A \\
2000 &  0.05 &  6.4  & 1.9 &  56  &  6.1  & 2.5 & 55 \\
 500 & all$^b$ & 1.6 & 1.6 &  6.7 & 1.6  & 1.5 & 2.8 \\
2000 & all$^b$ & 7.3 & 2.3 &  68  & 8.8  & 2.1 & 58 \\
\tableline
\end{tabular}
\\
$^{a}$~Mean of the best fit values for filaments fit at all times, with a non-zero
	background allowed for the fit and the central flat radius, $R_{fl}$ fixed
	at 0.01~pc: the
	central density, $\rho_c$ (in units of $10^5$~cm$^{-3}$),
	the exponent, $p$, and $\Sigma_0$, the background column density term 
	(in units of 10$^{21}$~cm$^{-2}$); the standard deviation is usually slightly
	less than the mean, with values of 0.9--2.3, and 
	0.2--0.7 for $\rho_c$ and $p$ respectively, in the same units.
	The background column density tends to have higher standard deviations
	than the mean values, and the mean is often much higher than the median
	value.\\
$^{b}$~Values of all profiles where a fit was possible are included here,
        i.e., relaxing the requirement of a fit for both HD and MHD,
        and for the 500~\Msol\ simulations, a fit at all time steps.
\end{table}

Although the variation between
filaments in a single simulation snapshot is larger than any general change in
filament behaviour as a function of time or initial conditions, the model
fits are consistent with the general behaviours noted earlier.  
Magnetic fields produce
puffier (lower $\rho_c$ and higher $R_{fl}$) filaments, and 
an initially higher density tends to lead to more peaky filaments (higher $\rho_c$ and 
lower $R_{fl}$); all of these differences are smaller than the scatter in best fit 
values for a given simulation snapshot, and would require tracking over a longer
time period to better measure time evolution.

As \citet{Arzoumanian11} found, we also 
find that the isothermal cylinder model tends to provide a worse fit to the
radial column density profile than the modified isothermal model, 
although this is much more noticeable 
in the case of a zero background column density (see Appendix~A).  
Typically, the best fit power law index, $p$ is between 1.3 and 2.0, 
within the range found by \citet{Arzoumanian11} and others, and much 
less than $p=4$ for the pure isothermal model. 
We also find $\rho_{c}$ is around $10^4$ to $10^7$~cm$^{-3}$,
and $R_{fl}$ ranges between roughly 0.001 to 0.1~pc, consistent
with \citet{Juvela12}.  The relationship
between $R_{fl}$ and the FWHM is dependent on the power law of the profile,
and that a smaller value of $R_{fl}$ is expected based on the values of $p$ fitted.
Explicitly, $FWHM = 2\times \sqrt{2^{2/(p-1)} -1} R_{fl}$, or 
$3.46 R_{fl}$ when $p=2$.

\subsection{Pressure-Confined Isothermal Cylinder Model Fits}
The pressure confined isothermal cylinder model also generally
provides a good fit to the filament profiles, although the HD example shown
in Figure~\ref{fig_radialfits_bkgrd} does a poor job of capturing the
peak column density.  In most cases, the best-fit model is very 
similar to the best-fit modified isothermal cylinder model.
The mean and standard deviation temperature of all fits 
are $15 \pm 14$~K, with a small tail in the temperature distribution extending
out to 100~K; only 20\% of the temperatures fit were above 20~K; the median
temperature fit is 11~K.
The typical external pressure fit was $P_{ext}/k_B = 4 \pm 3 \times 10^5$~cm$^3$~K$^{-1}$ ,
with the tail in the distribution extending up to $10^6$~cm$^3$~K$^{-1}$.  The
typical shape parameters fit were $f = 0.76 \pm 0.18$ (keeping in mind $f$ must be
between 0 and 1).  The fitted temperature and external pressure values are 
physically reasonable -- the temperatures are generally similar to the simulation's
constant 10~K, and the typical external pressure is is the same range as those
fitted and estimated to be reasonable in nearby molecular clouds such as 
Perseus \citep{Kirk06}.

We searched for evolutionary trends within the fitted model parameters, but
did not find any over the time period analyzed.  
The only discernable trend was that the external pressures fit tended
to be higher in the 2000~\Msol\ simulations 
than the 500~\Msol\ simulations, with typical values of $6 \pm 6 \time 10^5$~cm$^3$~K$^{-1}$ and
$1 \pm 2 \times 10^5$~cm$^3$~K$^{-1}$ respectively.  Since the initial density of the
2000~\Msol\ simulation was higher, the external pressure caused by the weight of
overlying material within the region would be expected to be higher.

Finally, we made a general comparison of the goodness-of-fit of the various models,
by comparing the typical (mean and standard deviation) chi-squared values of all fits.  The 
standard deviation in chi-squared values is several times larger than the mean for
any of the models fit, making a distinction in the overall goodness-of-fit between the models
tenuous.  Nevertheless, including a background column density
term for the isothermal cylinder model made a substantial difference: the mean
chi-squared value for fits with a background included is more than 3.5 times smaller
than when the background is zero.  The difference is much less pronounced for the
modified isothermal model where including a background column density
decreases the mean chi-squared value by 40\%.  Allowing the power law to vary
(purely isothermal model versus modified isothermal model) yields a 30\% improvement
in the mean chi-squared value, while the pressure confined model
has a mean chi-squared value 8\% lower than the modified isothermal model.

\section{Effect of Resolution}
Finally, we examine the effect of resolution on our results.  Real
observations of filamentary structures are complicated by both 
instrumental effects (including resolution and system noise) and
physical effects (how the flux emitted at a specific wavelength relates
to the intrinsic column density of material), which make direct
comparisons between simulations and observations difficult
\citep[e.g.,][]{Goodman09}.
Here, we consider only the simplest factor, spatial resolution,
to represent what `perfect' observations would be able to reveal.
For the analysis presented here, we assume that the observed system
is located 140~pc away, representing the very nearest molecular
clouds and therefore also a best-case scenario.  While we tested
a variety of resolutions, we show results from three cases which are
representative of the present single-dish facilities able to map
large areas of the sky in a reasonable amount of time.  The first
case we consider is a resolution of 8\arcsec.  This corresponds to
the resolution of the JCMT at 450~$\mu$m \citep{Holland13},
and is also similar to the 7\arcsec\ resolution of the recent CLASSY
survey \citep[e.g.,][]{FernLop14}
which studies in detail three nearby molecular clouds in unprecedented detail
with the CARMA interferometer.  The second case we consider is a 
resolution of 18.2\arcsec\, corresponding to the resolution of {\it Herschel}
column density maps using the group's latest standard method \citep[e.g.,][]{Palmeirim13},
with the resolution corresponding to that of their 250~$\mu$m observations.
The third and final case we consider is a resolution of 36.9\arcsec\ corresponding
to the resolution of {\it Herschel} at 500~$\mu$m, and earlier {\it Herschel}
column density maps \citep[e.g.][]{Konyves10,Arzoumanian11}.

\begin{figure*}[htb]
\includegraphics[width=7in]{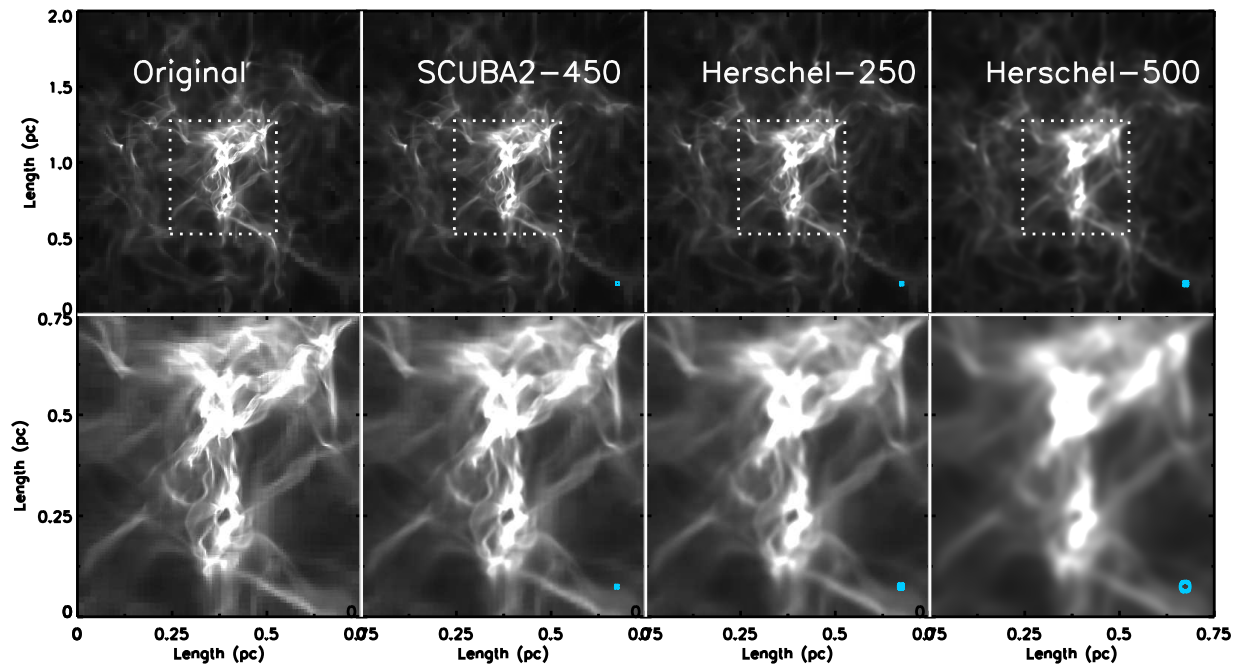}
\caption{A comparison of the column density in the original simulation and
	ideal observations at 8.0\arcsec, 18.2\arcsec, and 36.9\arcsec\ at 140~pc, 
	corresponding to SCUBA2 at 450~$\mu$m (or CLASSY) and {\it Herschel} at
	250~$\mu$m and 500~$\mu$m respectively (left to right).  
	The top panel shows the full 2~pc simulation
	box for the pure HD simulation with 500~\Msol\ viewed along the x axis
	at 0.15~Myr.  The bottom panel shows a zoomed in view to the box indicated in
	the top panel.  The blue circles indicate the beamsize / resolution for
	each panel.}  
\label{fig_col_smo}
\end{figure*}

Figure~\ref{fig_col_smo} shows an example of the effect of a resolution on 
the column density in the simulation.
As is clearly illustrated by this comparison, at the longest wavelength {\it Herschel} 
resolutions, much of the fine filamentary structure is lost, while SCUBA2 at 450~$\mu$m / CLASSY
does a better job.

We also examined the {\it quantitative} effect of the resolution on our results.  To
do this, we applied the same filament definitions and re-ran our analysis.
We attempted to correct for the resolution in a similar manner to real observational
analyses: for the direct filament FWHM measurements, 
we deconvolved the values with the resolution.
For the column density profile models, we convolved the model with the 
`beam profile' / resolution 
before fitting.  We find that despite attempting to correct for the resolution during
analysis using the standard observational techniques, 
the poorest resolution still gives biased results.
A full comparison of the effect of resolution on all of our measured quantities
is given in Table~\ref{tab_res}: we calculated the mean and standard deviation of
the ratio in values between the decreased resolution and original results.  In
summary, while there is a significant amount of scatter, typically the resolution
only produces a somewhat noticeable effect for the 36.9\arcsec\ or Herschel 500~$\mu$m
case.  The effects tend to be in the direction expected: poorer resolution leads to
higher widths and lower central densities.  We emphasize that these tests correspond to
the best possible case for these instruments.  Very few star-forming clouds are as
close as 140~pc; even Perseus is nearly twice as far away at $\sim$250~pc, and many
other Gould Belt Survey clouds are at a similar or larger distance.  Filaments surveyed
in the {\it Herschel} Hi-Gal Survey are several times more distant still.  
We ran additional tests (not shown) with even poorer spatial resolution, corresponding 
to filaments at these further distances, and found that the biases in observed
quantities becomes much more noticeable in those cases.

\citet{Smith14} tested the effect of resolution on measured filament widths
and find relatively little effect.  Their test used a degraded resolution of 0.0086~pc
(R. Smith, priv. comm.), corresponding to a resolution of $\sim$~13\arcsec\ at 140~pc.
\citet{Juvela12} also examined the effect of resolution on simulated filaments, using
an angular resolution of 40\arcsec\ at 93, 186, and 371~pc.  They found that at
the larger distances, the filament central densities were the most biased, while
the width, mass per unit length, and power law slope changed by less.
Both of these results are consistent with our findings at similar resolutions.

\begin{table}
\tablenum{6}
\centering
\tabletypesize{\scriptsize}
\caption{Effect of Resolution on Fit Parameters$^{a}$ \label{tab_res}}
\begin{tabular}{cccc}
\tableline
Resolution          & FWHM$^b$    & $\sigma^b$  & $M_{line}^b$ \\
\tableline
JCMT-450~$\mu$m     & $1.2\pm0.4$ & $1.1\pm0.2$ & $1.1\pm0.2$ \\
Herschel-250~$\mu$m & $1.3\pm0.5$ & $1.1\pm0.2$ & $1.2\pm0.3$ \\
Herschel-500~$\mu$m & $1.6\pm1.2$ & $1.3\pm0.4$ & $1.4\pm0.6$ \\
\tableline
Resolution	    & $\rho_c^c$  & $R_{fl}^c$  & $p^c$ \\
\tableline
JCMT-450~$\mu$m     & $0.8\pm0.3$ & $1.8\pm1.3$ & $1.2\pm0.4$ \\
Herschel-250~$\mu$m & $1.1\pm1.1$ & $1.5\pm1.1$ & $1.1\pm0.5$ \\
Herschel-500~$\mu$m & $0.5\pm0.3$ & $3.4\pm2.6$ & $1.6\pm0.7$ \\
\tableline
Resolution	    & $\rho_c^d$  & $R_{fl}^d$  & $p^d$ \\
\tableline
JCMT-450~$\mu$m     & $1.1\pm0.6$ & $1.1\pm0.3$ & -- \\ 
Herschel-250~$\mu$m & $1.1\pm0.6$ & $1.1\pm0.3$ & -- \\ 
Herschel-500~$\mu$m & $0.8\pm0.4$ & $1.5\pm0.7$ & -- \\ 
\tableline
Resolution 	    & $T^e$       & $P_{ext}^e$ & $f_{cyl}^e$ \\
\tableline
JCMT-450~$\mu$m     & $1.1\pm0.2$ & $1.2\pm0.5$ & $1.0\pm0.2$ \\
Herschel-250~$\mu$m & $1.2\pm0.6$ & $1.3\pm0.6$ & $1.0\pm0.2$ \\
Herschel-500~$\mu$m & $1.4\pm0.7$ & $1.4\pm0.8$ & $1.0\pm0.3$ \\
\tableline

\end{tabular}
\\
$^{a}$~Mean and standard deviation in the ratio between decreased resolution fits and original values (all quantities).\\
$^{b}$~Filament FWHM widths, Gaussian-fitted widths, and mass per unit length ratio.\\
$^{c}$~Parameters from the modified isothermal cylinder model.\\
$^{d}$~Parameters from the isothermal cylinder model (power law exponent fixed).\\
$^{e}$~Parameters from the pressure-confined isothermal cylinder model.\\
\end{table}

Beyond the implications to the measured filament properties presented here is the
presence and characterization of substructure within the filaments.  Averaging
a radial column density profile along a filament hides much of the information
on smaller-scale structure within the filaments in the simulations --
see, for example, the comparison of 2D and 3D column density versus
density profiles of simulated filaments in \citet{Gomez14} and \citet{Smith14}.
Both observations
\citep{Hacar13,Henshaw14,FernLop14}, and simulations \citep{Moeckel14,Smith14}
suggest that filaments may in fact be composed of multiple strands of dense gas,
perhaps woven together, which are often difficult to identify separately with the present 
observational capabilities; higher spatial resolution and / or inclusion of the line of
sight velocity are essential.  

In our hydrodynamic simulations too (see Figure~\ref{fig_col_smo}), 
we see evidence of more complex
structure on smaller scales, although a full 3D consideration of this is beyond
the scope of the present paper.  Nonetheless, our results coupled with new results
emerging from observations such as \citet{FernLop14} suggest that the key
to a deeper understanding of filamentary structure requires high (spatial and velocity)
resolution as well as more sophisticated tools by which to model the structure.

\section{Discussion \& Conclusions}
We simulate the formation of filaments within a clump-scale volume,
investigating the role of magnetic fields on the evolution
of filaments column density.  Starting with 500~\Msol\ (or 2000~\Msol)
of gas within a 2~pc cube, we track the evolution of filamentary structures over
0.15~Myr with and without the presence of a magnetic field.
Turbulence remains strong throughout the simulations as there has been been
insufficient time for it to damp significantly.

These analyses provide an important complement to the recent filamentary analysis
in simulations by \citet{Smith14}.  Those authors investigated the effect of different
initial modes of turbulence (e.g., solenoidal vs compressive), whereas our analysis
examines the effect of magnetic fields and gravity (through varying the initial
mean density).  Other more subtle differences are also important to note as well.  
\citet{Smith14} assumed a uniform density initial sphere, surrounded by a warm
diffuse medium, whereas we assume a sphere with a radially-decreasing density surrounded
by a vacuum.  \citet{Girichidis11} demonstrate that the initial density distribution can
have a marked effect on the large-scale structure which forms later in the simulation.
Our simulations do not include the effect of radiation or simple chemistry, while
\citet{Smith14} does include both; we expect these effects to become more important
at later times (once massive YSOs begin ionizing their natal environments) and when
making synthetic observations of molecular line emission (where the presence
or absence of various molecular species has a large effect).  The base numerical codes
used are also different - \citet{Smith14} use {\sc AREPO}, a hybrid code, while we
use \FLASH, which is an adaptive mesh refinement based code.  Despite these significant
differences, both \citet{Smith14} and our analyses do identify filaments which have
properties broadly similar to observed filaments, which appears to speak to the 
universality of filament formation under a variety of conditions.

We also note one major difference from \citet{Smith14} in the analysis stage: 
\citet{Smith14} focus their analysis on the brightest one or two filaments
in each simulation whereas we also include fainter / less dense filaments in
our analysis.  In this respect, our analysis gives a better direct comparison
with observational surveys, where many filaments are identified in any given
star-forming region.  Our approach limits our analysis of the filament column
density profiles to a smaller radial separation from the filament spine, since
the fainter filaments are more liable to have their profiles contaminated by nearby
non-related substructure than brighter filaments are (the filling factor is much
larger for all faint plus bright filaments than it is for just bright filaments).
On the other hand, inclusion of fainter filaments allows us greater sensitivity to
less gravitationally bound filaments, which may be more liable to re-expand with
time; such behaviour was not noted in \citet{Smith14}, presumably because the 
dominant filament in each simulation will continue to contract and accrete new
material throughout time.

Our main findings are as follows:
\begin{enumerate}

{\item {\it Magnetic fields have a strong influence on filamentary structure.}
Even with a mass to magnetic flux ratio which is
supercritical by a factor of $\sim$2, there are notable
differences from the purely hydrodynamic simulation.
Filaments formed in the magnetic case tend to be wider, less centrally peaked,
and evolve more slowly than filaments seen in the purely hydrodynamic
case. These differences are most apparent through a visual comparison
(e.g., Figures~\ref{fig_fils}, \ref{fig_radial1}, and \ref{fig_radial2}),
and are less discernable in quantitative measures due to the large variation
in filament properties at any given snapshot.}

{\item {\it The magnetic field can have a strong effect on the fragmentation of
filaments.}  In our simulations, magnetic fields are able to significantly suppress 
the formation of cores, since its energy density exceeds that of the turbulence.
The turbulence is sub-Alfvenic ($\mathcal{M}_A = v_A / \sigma \sim 2.1 - 2.2$)
and so the magnetic field increases the critical turbulent line mass
by a factor of 3.2 to 3.3.  
This accounts for the less condensed 
structure of the magnetized filaments, and their notably less fragmentation.}

{\item {\it The simulated filaments have properties consistent with observations.}
The radial column density profiles of the filaments are
well-described by a Plummer-like or modified isothermal cylinder profile.
The power-law slope tends to be around 1.3 - 2, similar to the range
of 1.5 - 2.5 found in {\it Herschel} data by\citet{Arzoumanian11}.
The central density tends to be of order $10^5$~cm$^{-3}$ for the 
500~\Msol\ simulations and closer to $10^6$~cm$^{-3}$ for the 2000~\Msol\
simulations; the inner flat radius is a few hundredths of a parsec
in both cases.  The 500~\Msol\ typical central densities, as
well as the inner flat radii and power law slopes are consistent with
those in \citet{Juvela12b}, also based on {\it Herschel} observations.
The pressure confined cylinder model of \citet{Fischera12}
{\it also} provides a reasonable fit to the radial column density profiles,
with typical temperatures, external pressures, and shape parameters fit
of 15~K, $4\times10^5$~cm$^3$~K$^{-1}$, and 0.76 for $T$, $P_{ext}/k_B$, and
$f$ respectively.
}

{\item {\it Filaments have diverse evolutionary paths.}
At any given snapshot in time, the simulation reveals a variety of 
filaments.  Some continue to radially contract and accrete material
throughout the simulation, and will presumably continue on to form
stars along their length.  Other filaments halt in their contraction
and expand into the ambient medium before the end of the simulation.
Given the relatively short time duration of our simulations,
we expect that even more diverse evolutionary paths could be possible
throughout the lifetime of a molecular cloud.}

{\item {\it The mass per unit length of a filament in a given snapshot provides
only a weak discriminant between the contracting and expanding filaments.}
Over most of the range of mass per unit length values, filaments can be 
either contracting or expanding.  Above roughly 72~Msol~pc$^{-1}$, the
critical value for filaments supported equally by thermal and non-thermal
support, nearly all filaments are contracting.
Velocity and magnetic field information are clearly required to determine
the evolutionary state of a filament unambiguously.
}

{\item {\it Turbulence plays an important role in the mass per unit length of 
filaments.}  The filaments in which stars appear in our simulations have critical 
mass per unit lengths that are dominated by turbulent velocity dispersion and not 
just thermal values.   This arises in these simulations since they are much less 
than a free-fall time old, so that turbulence has not had the opportunity to damp 
significantly.  {\it Herschel} results indicate that the {\bf thermal}
critical value of $M^{crit}_{line}$ is a good diagnostic of star-forming
ability therefore suggests that turbulence is much weaker within these
filaments, perhaps because their natal clouds are more evolved.
}

{\item {\it Filament widths are mildly influenced by environment.}
Filaments tend to be narrower at later times, when formed in higher
density environments, or without the presence of magnetic fields,
but these trends are all weak, given the mixture of contracting and
expanding filaments at every snapshot in the simulations.  
Stronger trends in the evolution in filament properties as a function
of time might become apparent with a longer timescale for analysis,
especially if the filaments which re-expand become sufficiently diffuse
to become undetectable at later times.
We find the mean filament FWHM ranges from 0.06 to 0.26~pc across
our simulations at varying times, with most being around 0.1~pc to 0.15~pc\footnote{See
Section~3.1 for a discussion on the biases and uncertainties in absolute
filament widths in our analyses.}.
The {\it range} in filament FWHM values for any given simulation snapshot
is, however, larger than the range seen in the analyses of \citet{Arzoumanian11}. 
For filaments which are contracting, a combination 
of the decaying turbulence and gravity is likely responsible for
the evolution.
This may in part explain the relative constant filament widths discussed
in \citet{Arzoumanian11}, although resolution may also be an
important factor \citep{FernLop14}, and observational biases and measurement methods
could be playing a role \citep{Heitsch13a,Smith14}.
}

{\item  {\it Filaments have complex structures.}
The radial column density profiles of the filaments reveal a wealth
of sub-structure, particularly in the pure hydrodynamic simulation
where features tend to be sharper.  
Some of these substructures appear
suggestively like the intertwined filament bundles found by \citet{Hacar13}
(see for example the bottom left panel in Figure~\ref{fig_col_smo});
high-resolution observations of a suite of filaments will be necessary to
show how commonplace this phenomenon is.
Other simulations, including \citet{Smith14} and \citet{Moeckel14} also
find complex 3D filamentary substructure.
}

\end{enumerate}

Future work will include a 3D analysis of the filaments formed in these simulations,
including their velocity structure and accretion rates, and the
relationship with the magnetic field geometry.

\begin{figure}[htb]
\includegraphics[width=3.5in]{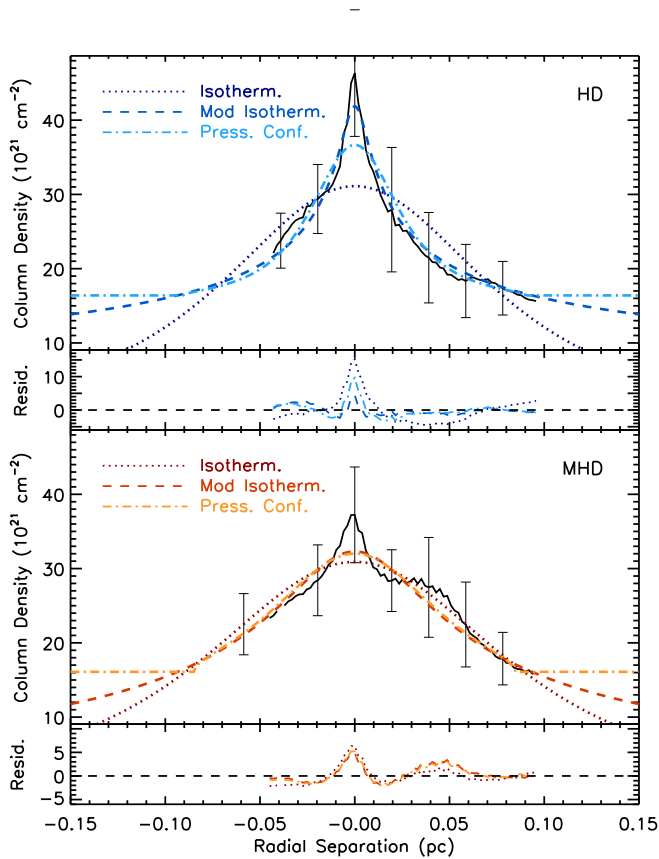}
\caption{The radial column density profile and best fit models for a filament, 
	including the fitting residuals where 
	the background column
	density level for the isothermal and modified isothermal cylinder models is
	fixed to zero.  
	See Figure~\ref{fig_radialfits_bkgrd} for the plotting conventions used.	
	Note that the vertical range in the residual plots differs between the
	upper and lower profiles.
	In this case, the pure isothermal cylinder model is not a good match to the profile.}
\label{fig_radialfits}
\end{figure}

\appendix
Here, we examine the results of fits to the radial column density profiles of the filaments
when the background column density is forced to be zero (see discussion in Section~4).
Figure~\ref{fig_radialfits} shows
the best fit models 
for the isothermal and modified isothermal model, for a filament identified
in both the HD (top panel) and MHD (bottom panel) 500~\Msol\ simulations,
to be contrasted with Figure~\ref{fig_radialfits_bkgrd} for the same fits where
background column density is allowed to be non-zero.  Note that the 
best-fit pressure confined model plotted in Figure~\ref{fig_radialfits}
{\it does} include a background column density term, and is identical
to the fit shown in Figure~\ref{fig_radialfits_bkgrd}.  Contrary to
the case (Section~4) when the background column density is fixed to zero, 
the three models differ from each other by a greater amount, and
the pure isothermal model is generally a poor fit to the profile.

\begin{table}
\tablenum{7}
\centering
\tabletypesize{\scriptsize}
\caption{Best fit parameters for modified isothermal cylinder, no background \label{tab_arz}}
\begin{tabular}{cccccccc}
Mass & 
Time &
\multicolumn{3}{c}{HD $^{a}$} &
\multicolumn{3}{c}{MHD $^{a}$} \\
(\Msol) &
(Myr) &
$\rho_c$ &
$R_{fl}$ &
$p$ &
$\rho_c$ &
$R_{fl}$ &
$p$ \\
\tableline
 500 &  0.05 &  1.1  & 1.7 & 1.4 &  0.7  & 2.0 & 1.4 \\
 500 &  0.10 &  3.1  & 0.7 & 1.4 &  0.7  & 1.8 & 1.4 \\
 500 &  0.15 &  3.9  & 0.6 & 1.5 &  2.3  & 1.2 & 1.4 \\
 500 &  0.20 &  2.6  & 0.8 & 1.5 &  N/A  & N/A & N/A \\
2000 &  0.05 &  11.1 & 0.7 & 1.3 &  7.9  & 0.6 & 1.2 \\
 500 & all$^b$ & 2.4 & 0.7 & 1.4 &  1.0  & 1.6 & 1.4 \\
2000 & all$^b$ & 11.1 & 0.5 & 1.2 & 10.7 & 0.6 & 1.3 \\
\tableline
\end{tabular}
\\
$^{a}$~Median of the best fit values for filaments fit at all times: the
	central density, $\rho_c$ (in units of $10^5$~cm$^{-3}$),
	the central flat radius, $R_{fl}$ (in units of 0.01~pc), and
	the exponent, $p$; the standard deviation is often comparable
	in magnitude to the median, with values of 0.7-32, 0.7-4.4, and 
	0.2-0.5 respectively, in the same units.\\
$^{b}$~Values of all profiles where a fit was possible are included here,
	i.e., relaxing the requirement of a fit for both HD and MHD,
	and for the 500~\Msol\ simulations, a fit at all time steps.
\end{table}

Table~\ref{tab_arz} gives the median model fit parameters
for the isothermal and modified isothermal profiles for all filaments fitted at 
every time step in both the HD and MHD
simulations. 
Comparison of these fit values with those given in Table~\ref{tab_arz_bkgrd}
shows that the best fit profiles tend
to have narrower peaks (better matching the filament profiles) when a background column
density included, most of this increased narrowness is accounted for by the steeper
power law, $p$, rather than a decrease in the central flat radius, $R_{fl}$.
As discussed in Section~4.2, there is relatively little difference in the quality of
fit (comparing typical $\chi^2$ values) for the modified isothermal model when
a background column density is or is not included, however, the inclusion of 
a background column density terms makes a substantial difference in the quality of
fits for the purely isothermal model.

\section{Acknowledgements}

We thank the referee for a thoughtful and thorough review which helped to strengthen our paper.
MK thanks Philip Girichidis for sharing his turbulence generator with us,
and Thierry Sousbie for providing us with a pre-release version of 
{\sc DisPerSE}.  HK thanks Rowan Smith for helpful discussions about
analyzing filaments in numerical simulations, especially in comparing her
recent results with ours.  HK thanks Doris Arzoumanian
for providing some clarification to the radial column density profile fitting
method used by the {\it Herschel} team.  HK also thanks Doug Johnstone and Phil
Myers for useful discussions about various aspects of this work.  HK acknowledges support
from the Banting Postdoctoral Fellowships program, administered
by the Government of Canada.  
MK acknowledges financial support from the National Sciences and Engineering
Research Council (NSERC) of Canada through a Postgraduate Scholarship.
REP is supported by a Discovery Grant from the Natural Sciences and Engineering Research 
Council (NSERC) of Canada. The \FLASH\ code was in part developed by the DOE NNSA-ASC OASCR 
Flash Center at the University of Chicago. This work was made possible by the facilities 
of the Shared Hierarchical Academic Research Computing Network (SHARCNET: www.sharcnet.ca) 
and Compute/Calcul Canada.

\bibliographystyle{apj}
\bibliography{ms}

\end{document}